\documentclass[a4paper,11pt]{article}
\pdfoutput=1 
\usepackage{aas_macros}
\usepackage{jcappub} 
\usepackage[T1]{fontenc} 
\usepackage{amsbsy}
\usepackage{color}
\usepackage{xcolor}
\usepackage{subcaption}
\usepackage{hyperref}

\usepackage[normalem]{ulem}

\newcommand{\be}{\begin{equation}}
\newcommand{\ee}{\end{equation}}

\usepackage{amssymb}

\title{\boldmath Exploring the effects of primordial non-Gaussianity at galactic scales}

\author[a]{Cl\'ement Stahl}
\author[b]{Thomas Montandon}
\author[a]{Benoit Famaey}
\author[b,c]{Oliver Hahn}
\author[a]{Rodrigo Ibata}

\affiliation[a]{Universit\'e de Strasbourg, CNRS, Observatoire astronomique de Strasbourg, UMR 7550, 67000 Strasbourg, France}
\affiliation[b]{Department of Astrophysics, University of Vienna, T\"urkenschanzstra{\ss}e 17, 1180 Vienna, Austria}
\affiliation[c]{Department of Mathematics, University of Vienna, Oskar-Morgenstern-Platz 1, 1090 Vienna, Austria}

\emailAdd{clement.stahl@unistra.fr}
\emailAdd{thomas.montandon@univie.ac.at}

\abstract{While large scale primordial non-Gaussianity is strongly constrained by present-day data, there are no such constraints at Mpc scales. Here we investigate the effect of significant small-scale primordial non-Gaussianity on structure formation and the galaxy formation process with collisionless simulations: specifically, we explore four different types of non-Gaussianities. All of these prescriptions lead to a distinct and potentially detectable feature in the matter power spectrum around the non-linear scale. The feature might have interesting consequences for the $S_8$ tension. We then show in particular that a negatively-skewed distribution of the potential random field, hence positively skewed in terms of overdensities, with $f_{\rm NL}$ of the order of 1000 at these scales, implies that typical galaxy-sized halos reach half of their present-day mass at an earlier stage and have a quieter merging history at $z<3$ than in the Gaussian case. Their environment between 0.5 and 4 virial radii at $z=0$ is less dense than in the Gaussian case. This quieter history and less dense environment has potentially interesting consequences in terms of the formation of bulges and bars. Moreover, we show that the two most massive subhalos around their host tend to display an interesting anti-correlation of velocities, indicative of kinematic coherence. All these hints will need to be statistically confirmed in larger-box simulations with scale-dependent non-Gaussian initial conditions, followed by hydrodynamical zoom-in simulations to explore the detailed consequences of small-scale non-Gaussianities on galaxy formation.}

\begin{document}
\maketitle
\flushbottom

\section{Introduction}\label{sec:Introduction}
 \subsection{Challenges for galaxy formation simulations}
 \label{sec:Simvsgal}
 
The standard cosmological model \cite{Ostriker, Planck:2018vyg} assumes that the matter content of the Universe, dominated by cold dark matter (CDM), formed structures out of primordial nearly-scale invariant Gaussian perturbations following the period of inflation. In this model, late-time acceleration is accounted for by a cosmological constant ($\Lambda$), and it is therefore dubbed $\Lambda$CDM. This model offers a framework to fit and interpret almost every large-scale observation of our Universe, although some large-scale tensions are still actively discussed \citep[e.g.,][]{Abdalla}. On smaller, galactic scales ($\sim \mathcal{O}(1)$ Mpc/h), observational challenges to $\Lambda$CDM have also been put forward for the last two decades. They have remained actively discussed \citep[e.g.,][]{Famaey, Bullock, Peebles2022} whilst both galactic-scale observations and numerical simulations got better and more precise \citep[e.g.,][]{Auriga,Fire,Pillepich,Vogelsberger:2019ynw,Dubois,Angulo:2021kes}. Baryonic feedback is most often invoked to solve, or at least ease, many of the galactic-scale tensions \citep{Bullock}, but the jury is still out on how fine-tuned the feedback processes need to be for the solution to be natural. There are, however, challenges that feedback typically cannot address: on slightly larger scales than galaxies themselves, the observed planar configuration of satellite galaxies with correlated kinematics, both in the Local Group and beyond \citep{Pawlowski1, Ibata:2014csa, Muller}, remains mostly unexplained and should be largely independent, at least to first order, from feedback. 

Among all these challenges, one of the oldest problems of galaxy formation in the standard $\Lambda$CDM context was the cosmological angular momentum problem \citep[e.g.,][]{Katz91,Navarro91,DOnghia:2004iex}, namely that numerical simulations naturally tended to produce galaxy discs that were too small and much more centrally concentrated than is observed. This problem has largely been improved upon in galaxy formation simulations through the implementation of various feedback recipes, but, as pointed out by, e.g., Peebles \cite{Peebles:2020bph}, the most recent hydrodynamical simulations \citep{Auriga, Fire} of galaxy formation still suffer from a `hot orbits problem'. Comparing the fraction of stars belonging to spheroidal components (bulges and stellar halos) in $L_*$ galaxies within 10~Mpc (the median fraction being of 15\%) to those in modern galaxy formation simulations (that predict a typical fraction of 45\%) reveals a truly resilient tension: simulated galaxies are too bulgy or have too massive stellar haloes. This recent version of the angular momentum problem may indicate that Milky Way-like disk galaxies do not have a quiet enough history in the $\Lambda$CDM context. A problem that is probably related is the inability of simulations to form the right fraction of barred galaxies as a function of stellar mass \citep{Reddish, Roshan}, which also form bars that are too small for a given pattern speed \citep[][]{Roshan,Frankel}, although see also Ref.~\cite{Fragkoudi}. While the latter problem may be linked to numerical resolution issues, interestingly, bar formation naturally arises in idealized simulations of isolated disk galaxies, and is only inhibited in a cosmological context. Hence having a quieter and more underdense environment could allow cosmological simulations to resemble more closely idealized simulations and potentially alleviate this tension.  

While such challenges on small-scales might point to radical alternatives to $\Lambda$CDM, involving the nature of dark matter itself, another change that is worth exploring is to modify the initial conditions of structure formation. Some approaches developed in recent years include the concept of `genetic modification' \cite{Rey:2018cfb,Stopyra:2020egb} and `splicing' \cite{Cadiou:2021chs}, or changing the initial angular momentum distribution \cite{Cadiou:2022krq}. Another more \textit{ab initio} perspective, as the early universe is assumed to be Gaussian in $\Lambda$CDM, is to implement non-Gaussianities in the primordial density fluctuations on scales of a few Mpc.
The effect of large-scale non-Gaussianities has already been explored in previous simulations, with a focus on the matter power spectrum, halo mass function and halo density profiles \citep[e.g.,][]{Avila-Reese:2003cjm,Dalal:2007cu,LoVerde:2007ri,Pillepich:2008ka,Smith:2010fh,MoradinezhadDizgah:2013rkr}. Here, we rather concentrate on non-Gaussianities on small-scales, and trade large-box size for better resolution, which is well-suited to study the impact of such small-scale non-Gaussianities on typical galaxy-sized halo scales. As suggested by, e.g., Peebles \cite{Peebles:2020bph}, this could lead to a different, quieter environment for Milky Way-like galaxies and a different merger history that might allow for subtle changes in the disk and bar formation, as well as in the distribution of subhalos. This is the hypothesis we explore in the present paper.

As a first exploratory step in this direction, we present hereafter gravity-only collisionless simulations of structure formation with non-Gaussianities on small scales to investigate the potential variations with respect to ``vanilla'' $\Lambda$CDM. Our results indicate an influence of non-Gaussianity of the initial conditions on the assembly history of galaxy-sized halos. Future work will have to corroborate these preliminary results by including also baryonic physics and simulating the full formation of galaxies.

 \subsection{Non-Gaussianities}
 
The small-scale tensions mentioned above have motivated the community to investigate several potential solutions, ranging from alterations of the nature of dark matter to modifications of gravity. In contrast to earlier work, here we propose to explore the effect of changing the small-scale initial conditions of our Universe, thereby altering slightly the formation history of structures. This minimal change is not mutually exclusive with other modifications of the dark sector, but we choose to concentrate on these to isolate their potential effects. We do that by implementing small-scale primordial non-Gaussianity (PNG) for collisionless (gravity-only) simulations of structure formation in a traditional N-body code \citep{Springel:2020plp}.

It is important to remember that PNG are actually a natural prediction of {\it all} existing inflation models \cite{Maldacena:2002vr,Creminelli:2004yq,Biagetti:2019bnp}, albeit in general with a very low amplitude. Indeed, if a single degree of freedom was active during inflation, powerful theorems predict some small amount of PNG: the amplitude of PNG is labelled by the parameter $f_{\rm NL}=\frac{5}{12}(1-n_S)  \sim 0.01$ \cite{Maldacena:2002vr} where $n_S$ is the spectral tilt of the power spectrum (see next Section for formal definitions in eqs 2.1--2.4). The parameter $f_{\rm NL}$ is typically proportional to the slow-roll parameters. More elaborate models of PNG can however predict a much larger amount of PNG \cite{Biagetti:2019bnp}. Much effort has been dedicated to put stringent observational constraints on PNG and on the $f_{\rm NL}$ parameter \cite{Planck:2019kim,Mueller:2021tqa,Cabass:2022ymb} from the Cosmic Microwave Background (CMB) and Large Scale Structure (LSS). At smaller scales ($\sim \mathcal{O}(1)$ Mpc/h), the effects of PNG have not been extensively explored (see however Ref.~\cite{Chevallard:2014sxa} for an attempt) and are difficult to test observationally. 

If inflation happened as a roughly scale invariant process controlled by the slow roll parameters, the constraints imposed by large scales measurements would also hold at smaller scales. But some models beyond slow roll do allow for scale-dependent non-Gaussianities that could be large at galactic scales \cite{Khoury:2008wj,Riotto:2010nh,Byrnes:2011gh}. Tentative constraints on such models have been attempted using LSS data and galaxy luminosity functions (eg.~\cite{Khatri:2015tla,Sabti:2020ser,Rotti:2022lvy,Bianchini:2022dqh}). Of particular relevance for our exploration, the UV galaxy luminosity function derived from the Hubble Space Telescope observations provided a tentative constraint showing that when PNG are only present at scales smaller than $\sim 6$~Mpc, their best fit becomes $f_{\rm NL} \approx -1000$, with a departure from $f_{\rm NL} =0$ significant at $1.7 \sigma$ \citep{Sabti:2020ser}.

In this work, we assume that such an \textit{effective} local non-Gaussian signal is only present on small scales ($\sim \mathcal{O}(10)$ Mpc/h) thus mostly evading the CMB constraints. We leave for future work the realisation of such a scenario in primordial physics (for instance the link with the formalism of Ref.~\cite{Riotto:2010nh}). We indeed focus on a bottom-up approach to the development of small-scale structures and the (possible) resolution of (some of the) aforementioned small-scale problems of $\Lambda$CDM.

In section \ref{sec:NGIC}, we present how we modified the initial conditions of the N-body code to take into account small-scale non-Gaussianity with four different models. We also review in that section the existing observational constraints on PNG. In section \ref{sec:LSSdevelopement}, we heuristically describe the expected main differences in the development of structures in these non-Gaussian scenarios. In section \ref{sec:resul}, we present the different quantities that we extracted from our N-body experiments: merger trees, density of the environment at $z=0$, initial and final power spectra, halo mass function at $z=0$, and a measure of the kinematic coherence of satellites around host galaxies. Finally, section \ref{sec:concl} presents the conclusions of the study, in which we draw some perspectives and wrap up the discussion. 

 \section{Non-Gaussian initial conditions} 
 \label{sec:NGIC}

 \subsection{Non-Gaussian templates}\label{sub:template}
 
 Primordial fluctuations in a statistically homogeneous and isotropic Universe are described as a random field, best described in Fourier space: the variance of the modulus of the Fourier transform $|\delta(\mathbf{k})|$ of the contrast density field $\delta(\mathbf{x}) = \frac{\rho(\mathbf{x})-\rho_0}{\rho_0}$ (where $\rho_0$ is the background density) is provided by the Fourier transform of the two-point correlation function $\xi(r)$, namely the power spectrum $P(k) \propto \langle |\delta(\mathbf{k})|^2 \rangle_{|\mathbf{k}|=k}$.
 
 For a Gaussian random field, where the contrast density field $\delta(\mathbf{x})$ is Gaussian and $|\delta(\mathbf{k})|$ has a $\chi$-distribution with 2 degrees of freedom (and assuming uniformly-distributed phases), all the information is contained in the two-point correlation function. Exactly the same reasoning applies to the gravitational potential generated by the contrast density field. Hence the information in a Gaussian random field $\Phi_G$ for the gravitational potential is fully described by its two-point correlation function
 \be
 \langle \Phi_G(\mathbf{k}_1) \Phi_G(\mathbf{k}_2) \rangle = (2\pi)^3 \delta_D(\mathbf{k}_1+\mathbf{k}_2) P_{\Phi}(k_1) \, , \label{eq:2ptcorr}
 \ee
where $\delta_D$ is the Dirac delta and, in cosmology, the primordial power spectrum $P_{\Phi}$ has the form
\be
\label{eq:P_Ini}
P_{\Phi}= \frac{2 \pi^2}{k^3} \frac{9}{25} A_s \left(\frac{k}{k_{\rm pivot}} \right)^{n_S-1}.
\ee
The parameters $n_S$ and $A_s$ are the difference to scale invariance and the amplitude of primordial perturbation, while $k_{\rm pivot}=0.05 \text{ Mpc}^{-1}$ is some fixed momentum and the factor $9/25$ comes from the relation between the gravitational potential $\Phi$ and the primordial potential curvature perturbation $\zeta(x)$, namely $\Phi = -3/5 \zeta(x)$ \cite{Maldacena:2002vr}.
 
For a non-Gaussian random field $\Phi$, the three-point correlation function and its associated bispectrum $B_\Phi$ become non-zero: 
 \be
 \langle \Phi(\mathbf{k}_1) \Phi(\mathbf{k}_2)\Phi(\mathbf{k}_3)\rangle=(2\pi)^3 \delta_D(\mathbf{k}_1+\mathbf{k}_2+\mathbf{k}_3) B_{\Phi}(k_1, k_2,k_3)\,.
 \ee
 In cosmology, such non-Gaussian random fields are traditionally expanded around their Gaussian values with $f_{\rm NL}$ quantifying the amount of PNG associated with the three-point correlation function and $g_{\rm NL}$ being the equivalent for the four-point correlation function \footnote{ Note the sign $-f_{\rm NL}$ in the expansion when it is expressed in terms of the gravitational potential related to density fluctuations via Poisson's equation, implying that a positive $f_{\rm NL}$ will be positively skewed in terms of overdensities as per the traditional convention.}

\be
\label{eq:PNGtemplate}
 \Phi(\mathbf x)=\Phi_G(\mathbf x) - f_\mathrm{NL} \left( \Phi_G^2(\mathbf x)  - \langle \Phi_G^2 \rangle \right) + g_\mathrm{NL} \Phi_G^3(\mathbf x)\,.
\ee

The Planck satellite gave exquisite measurements of the power spectrum and bispectrum confirming that the universe is Gaussian on large scale \cite{Planck:2019kim} with bispectrum measurements compatible with zero. These measurements were complemented by LSS probes establishing the vanilla $\Lambda$CDM paradigm, see eg.~\cite{Cabass:2022ymb} for a measure of $f_{\rm NL}$. These constraints are however not valid on small scales for scale-dependent non-Gaussianities.

\begin{table*}
\begin{center}
\caption{Summary of the different simulations carried out in this paper. The columns give the simulation identifier and the values of the PNG parameters $f_1$, $f_2$, $f_{\rm NL}$ and $A$. 
}
 \begin{tabular}{ || l || c | c | c | c |  c || }
\hline\hline  
Simulations & G &  NG1+ &  NG1- & NG2+ & NG2-  \\
\hline
$f_{1}$ & 0 & 0.1 & -0.1 & 0 & 0  \\
$f_{\rm NL}$ & 0 & -1031 & 959 & 0 & 0  \\
$f_{2}$ & 0 & 0& 0& 0.1 & -0.1 \\
$A$ & 1 & 1.04 & 0.96 & 0.77 & 1.44 \\
\hline \hline
\end{tabular}
\label{table:simtable}
\vspace{-5mm}
\end{center}
\end{table*}

\begin{figure}
\centering
    \includegraphics[width=\textwidth]{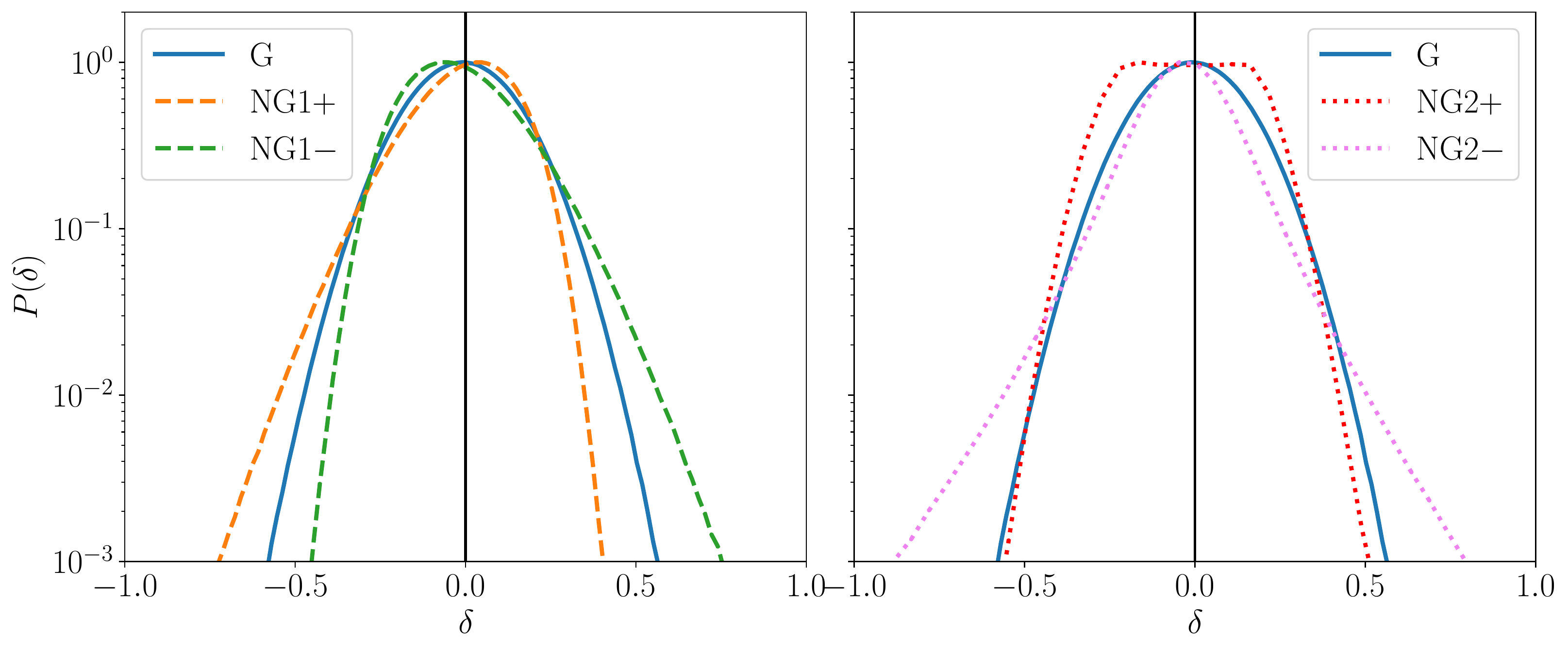}
    \caption{Histogram of the measure of the initial ($z=50$) density contrasts $\delta$ generated with Equations \eqref{eq:fnl?} and \eqref{eq:gnl?}. Each density contrast is computed on a grid cell of comoving length 58~kpc/$h$. These correspond to typical non-Gaussian PDFs with skewness (NG1$\pm$) and kurtosis (NG2$\pm$).}
 \label{fig:hist}
\end{figure}

In Ref.~\cite{Peebles:2020bph}, the author proposes to adopt a local expansion of the density up to third-order at the initial time of the simulation for small-scale non-Gaussianities
 \be
 \label{eq:fnl?}
 \Phi(\mathbf x)=A \left(\Phi_G(\mathbf x) + \frac{f_1}{\langle \Phi_G^2 \rangle^{1/2}} ( \Phi_G^2(\mathbf x)  - \langle \Phi_G^2 \rangle) \right),
 \ee
  \be
 \label{eq:gnl?}
  \Phi(\mathbf x)=A \left(\Phi_G(\mathbf x) + \frac{f_2}{\langle \Phi_G^2 \rangle} \Phi_G^3( \mathbf x) \right) \,.
 \ee
In Fig.~\ref{fig:hist}, we provide a graphical representation of such templates in terms of the distribution of associated contrast densities $\delta$. The normalisation factor $A$ is chosen such that a (linear) measure of $\sigma_8$ would coincide in the Gaussian and non-Gaussian cases. The cosmological parameter $\sigma_8$ is defined as the standard deviation of the amplitude of fluctuations when sampling the Universe at random places within spherical volumes of 8$h^{-1}$Mpc, corresponding to a scale that is still in the linear regime of cosmological perturbations. In practice, to ensure a common $\sigma_8$, we first generated our non-Gaussian template with $A=1$, we then measured $\sigma_8^{NG}$ and subsequently rescaled our ICs by fixing $A=\sigma_8^{NG}/\sigma_8$. The values that we have found for our specific numerical setups are reported in Table \ref{table:simtable}. We have performed this procedure in order to ensure that the results that we present in this work are due to the change of shape of the Probability Distribution Function (PDF) and not simply to an increase or decrease of the power on all scales, this will become relevant when discussing the $S_8$ tension. The relation between our parametrization and the usual $f_{\rm NL}$ and $g_{\rm NL}$ is
 \be
 f_{\rm NL}= - A \frac{f_1}{\langle \Phi_G^2 \rangle^{1/2}},
 \ee
  \be
 g_{\rm NL}=A \frac{f_2}{\langle \Phi_G^2 \rangle}.
 \ee
We obtain our initial linear density by (linearly) propagating the different templates with the Boltzmann code \texttt{CLASS} \cite{Blas:2011rf}. We define the transfer functions $T(t,k)$ which links the density contrast at the cosmic time $t$ to the gravitational potential:
 \be
\label{eq:transfer}
  \delta(t,\mathbf k) = T(t,k) \Phi(\mathbf k) \,.
\ee

In Fig.~\ref{fig:hist}, we display the histograms of the initial conditions (ICs) density perturbations for the four different cases we explored in this paper, all summarised in Table~\ref{table:simtable}.  For $P(\delta)>10^{-3}$, these are a faithful measurement of the different PDFs. The blue curve is the Gaussian case. In the left panel, the dashed orange is a skewed model with a positive non-Gaussian part $f_1$, dubbed NG1+ (hence with a negative $f_{\rm NL}$), which adds an asymmetric component. Since the density transfer function $T(\eta,k)$ is negative, a positive $f_1$ induces a skewness towards larger underdensities, which translates into a larger tail for negative $\delta$. On the other hand, the over-densities are less likely and smaller. The mode of the distribution (most likely perturbation) is on the other hand a small over-density. The case $f_1<0$ is the parity symmetric of the dashed red curve, which we dub NG1-, with a skewness towards large overdensities  (positive $f_{\rm NL}$) and the mode of the distribution peaking at a slight underdensity. In the right panel, we consider cases with $f_2 \neq 0$, such that the non-Gaussian term is symmetric with respect to zero. The case $f_2>0$, which we dub NG2+, has a single bump centred on $0$ and larger tails, i.e. larger over- and under-densities in the ICs. However, the density perturbations of small amplitude, e.g. around $|\delta| \sim 0.4$, are less likely in that case. The case $f_2<0$, dubbed NG2- in dotted magenta, is characterised by two bumps around $0$. This bimodality is an artefact of having a large $g_{\rm NL}$ while all higher coefficients are kept at zero. We explore this model to isolate the effects of the $g_{\rm NL}$ coefficient although, in reality, inflation would never produce such a non-Gaussian distribution for which only one higher moment is non-zero. In this (artificial) case, the most likely perturbations are around $|\delta|\sim 0.1$. Note that if one wishes to increase the amount of overdensities with $\delta \sim 0.5$ and above with respect to the Gaussian case, the two scenarios of relevance are NG1- and NG2+.

\subsection{Selected observational constraints on non-Gaussianities}

\label{sec:constrain_fnl}

The best constraints on large-scale primordial non-Gaussianities have been obtained with the analysis of the CMB temperature and polarisation bispectrum. In Ref.~\cite{Planck:2019kim}, the local shape is constrained with an amplitude $f_{\rm NL} = -2.5 \pm 5.0$ at $1 \sigma$. This constraint was obtained with the binned bispectrum estimator \cite{Bucher:2015ura} for the multipoles $\ell \in \left[2  ; 2500 \right]$ for the temperature and $\ell \in \left[ 4 ; 2000 \right]$ for the polarisation. The constraints on the trispectrum gave $g_{\mathrm{NL}} = (-5.8 \pm 6.5) \times 10^{4}$. The highest multipole $\ell_{\rm max}$ can be linked with the comoving size $d_{\rm min}$ 
\be
 \label{eq:lmax}
 d_{\rm min} = \frac{\pi}{\ell_{\rm {max}}} D_M
\ee
where $D_M$ is the angular diameter distance of the CMB and $z_*$ its redshift. By using the constraints on the acoustic angular scale $\theta_*$ and the comoving sound horizon at recombination $r_*$ given in Ref.~\cite{Planck:2018vyg}, we find $D_M \approx 14$\,Gpc. Hence the smallest comoving scale where the non-Gaussianities have been probed by the CMB is $d_{\rm min} \sim 11\, h^{-1}$\,Mpc. 

Concerning LSS constraints, Refs.~\cite{Mueller:2021tqa,Rezaie:2021voi}, using the eBOSS quasar clustering, found $f_{\rm NL} = -12 \pm 21$ at $1 \sigma$ using a data range $k \in \left[0.0019, 0.121 \right]$ $h$/Mpc, thus the smallest scales probed were $d_{\min}=52$ Mpc/$h$. In the same vein, in Ref.~\cite{Cabass:2022ymb}, the BOSS galaxy survey was analysed to put constraints on PNG: $f_{\rm NL} = -33 \pm 28$ at $1 \sigma$. The data cut used for the analysis was $k \in \left[0.01, 0.25 \right]$ h/Mpc leading to a minimal distance $d_{\min}=25$ Mpc/$h$. We note that all the best-fit values, both from LSS and CMB, although widely compatible with zero, are negative. Note also that all these studies assume that PNG is scale-invariant and local, meaning that the constraints on the scale of $d_{\rm min}$ {\it alone} should be weaker if one allows for scale-dependence.

On smaller scales, which our present study is concerned with, using the spectral $\mu$-distortions from Planck and correlating them with temperature fluctuations, Ref.~\cite{Rotti:2022lvy} have constrained a small scale $f_{\rm NL}<6800$ at 2 $\sigma$. As the $\mu$ distortions are sourced by dissipation damping at scales $\mathcal{O}(1)$ kpc/$h$, this allows one to probe non-Gaussianities at those scales. See also Ref.~\cite{Bianchini:2022dqh} for a similar study with the addition of the FIRAS data. Also, of particular relevance for our exploration, the UV galaxy luminosity function from the Hubble Space Telescope provided the constraint $f_{\rm NL} = 71^{+426}_{-237}$ at $2 \sigma$ \cite{Sabti:2020ser}, for PNG only present at scales smaller than 60~Mpc. However, when PNG are only present at scales smaller than 6~Mpc, the best fit becomes $f_{\rm NL} \approx -1000$, with a departure from $f_{\rm NL} =0$ significant at $1.7 \sigma$.

In this paper, in order to explore the effect that small-scale PNG could have on galaxy formation, we chose a simulation box with linear size of 30 Mpc/$h$. This is close to the value of $d_{\rm min}$ from large scale constraints by using the templates \eqref{eq:fnl?} with values corresponding to $f_{\rm NL} \approx \pm 1000$ (see table \ref{table:simtable} for the exact values), and large values of $g_{\rm NL}$ for templates \eqref{eq:gnl?}. However as those local templates are not scale-dependent, if we were to simulate larger scales with the same setup, we would not evade the Planck constraints. Hence, in this work $f_1$ and $f_2$ should be interpreted as effective amplitudes of a scale-dependent PNG template for the scales of interest, typically smaller than $d_{\rm min}$ in this article. With such a toy model, our goal is to qualitatively evaluate the impact of large-amplitude PNG on small scales. We leave for future work the elaboration of more realistic models with scale dependent non-Gaussianities and their potential link to inflationary physics.

 \subsection{Initial condition generator}
 \label{sec:MonofonIC}
We modified the \texttt{MonofonIC} software package\footnote{The official code is available at \url{https://bitbucket.org/ohahn/monofonic/} to include PNG in the  generation of initial conditions for simulations. Our modified branch can be found at \url{https://bitbucket.org/tomamtd/monofonic/}.} to generate the initial conditions, see Refs.~\cite{Hahn:2011uy,Michaux:2020yis}. The Gaussian primordial curvature perturbation $\zeta_G$ is generated by multiplying a white noise field $W(\mathbf k)$ by the square root of the primordial power spectrum $P_\zeta$ (given by multiplying Eq.~\eqref{eq:P_Ini} by 25/9):

\be
\label{eq:zeta_field}
\zeta_G(\mathbf k) = W(\mathbf k) \left( \frac{A_s }{4\pi k^3 } \right)^{1/2} \left( \frac{k}{ k_p } \right)^{(n_s-1)/2} \,.
\ee
This curvature perturbation can be used to compute the full primordial initial conditions from Eqs.~\eqref{eq:fnl?} and \eqref{eq:gnl?}. As noted, for instance in Ref.~\cite{Dalal:2007cu}, non-Gaussian templates feature multiplications of fields that lead to aliasing errors on large scales. These errors come from the periodic numerical representation of the field. In \texttt{MonofonIC}, aliasing is avoided by using Orszag’s $3/2$ rule, see Ref.~\cite{Michaux:2020yis} for more details. To our knowledge previous PNG simulations did not correct for aliasing due to PNG non-linearities. The  matter density at the initial redshift of the simulation can then finally be computed by using Eq.~\eqref{eq:transfer}.

 \begin{figure}
     \centering
     \includegraphics[width=\textwidth]{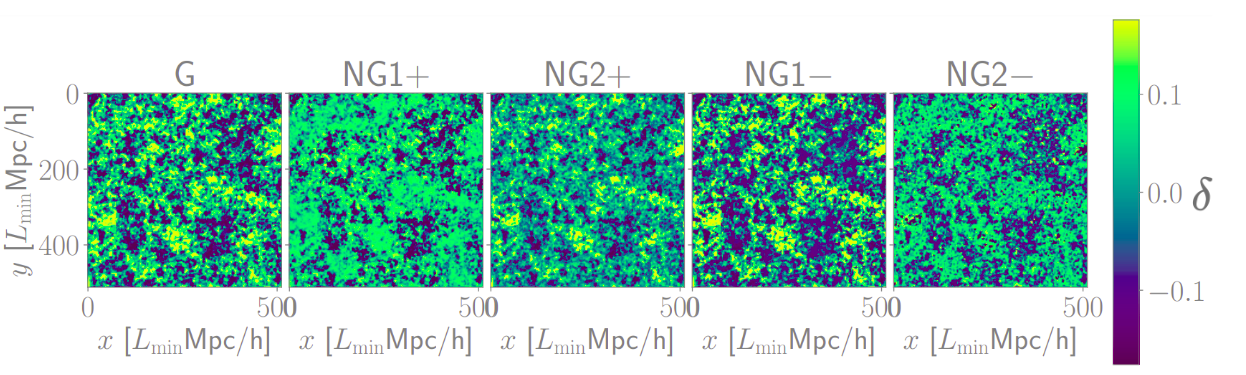}
     \caption{Density contrast slices for our ICs at redshift $50$ for the five cases under study, see Table \ref{table:simtable} for the nomenclature. The minimum comoving scale $L_{\rm min}$ is the size of the box divided by the number of grid points $L_{\rm min} = (30/512)$~Mpc/$h$. }
     \label{fig:IC_imshow}
 \end{figure}

In Fig.~\ref{fig:IC_imshow}, we show slices through the linear matter density field used by the modified \texttt{MonofonIC} to compute the displacement field at redshift $z=50$ on a $512^3$ grid. For $f_1>0$, we have seen in sect.~\ref{sub:template} that the peak of the PDF is shifted towards small overdensities: hence, on the second panel, we see this behaviour by remarking that the figure is more greenish, but it does not reach large overdensities in yellow. Most of the large perturbations are underdensities in blue. In the fourth panel, for $f_1 < 0$, this behaviour is exactly reversed, as expected, with large overdensities being surrounded by a slightly underdense peak of the distribution. The third and fifth panels correspond to $f_2\neq 0$. For $f_2 > 0$, the perturbations appear very similar to the Gaussian case, but they are actually sharper, with some larger overdensities. In the last panel, for $f_2 < 0$, we clearly see that most of the perturbations are coloured in hues of green and blue, which correspond to the two peaks discussed in sect.~\ref{sub:template}.

 \subsection{Numerical setup}
 \label{sec:NumSetup}
The numerical initial conditions described in section \ref{sec:MonofonIC} are then translated into particle perturbations using Lagrangian perturbation theory at first order. These are applied at our initial redshift $z=50$ into the Tree-PM code \texttt{Gadget 4}\footnote{\url{https://gitlab.mpcdf.mpg.de/vrs/gadget4}} \cite{Springel:2020plp} using a gravitational softening length of $\epsilon=0.5$ kpc.
For all simulations we adopt the following cosmological parameters: $\Omega_M=0.309899$, $\Omega_{\Lambda}=0.6901$, $h=0.67742$, $A_s=2.1064 \times 10^{-9}$ and $n_s=0.96822$. We have chosen a box length $L=30$ Mpc/$h$, in which we simulate a total mass of $3.4 \times 10^{15}\, M_{\odot}$. With $512^3 \approx 1.34 \times 10^8$ particles, our mass resolution is $2.6 \times 10^{7}\, M_{\odot}$.  Table \ref{table:simtable} sums up our nomenclature and the most relevant quantities for the different simulations carried out in the paper. 

An amplitude of $0.1$ for $f_1$ and $f_2$ implies that the non-Gaussian part increases or decreases by $10$ \% the Gaussian field, justifying that we can still write the model as an expansion over the Gaussian field. In terms of $f_{\rm NL}$, it would correspond to $f_{\rm NL}= \pm \mathcal{O}(1000)$ that is obviously ruled out on large scales but is still possible on smaller scales, as discussed in detail in section \ref{sec:constrain_fnl}.
 
\section{An invitation to revisit structure formation}
\label{sec:LSSdevelopement}

Before delving into the results of our simulations, we briefly explain here heuristically what would {\it a priori} be expected with the four different types of non-Gaussianity considered here in terms of small-scale structure formation. 

In particular, two models should have a tendency to form more large mass halos and to form them slightly earlier than in the Gaussian case: NG1- and NG2+. Because large overdensities would be present from the start in these models, one would expect them to have a quieter merging history at later times. In the NG1- case, the peak of the distribution of the  density contrast  is slightly shifted towards underdensities, which could imply that the environment of large-mass halos would be less crowded. This would also imply a quieter merging history, which in turn could have important consequences on the formation of bulgeless disks or bars. This quieter merging history could also have consequences on the phase-space distribution of subhalos around their host, as they could fall into the host virial radius in fewer group infalls than in the Gaussian case, and end up having slightly more coherent structure at the present day. 

We will test these expectations based on our simulations in the following section. For this, we will concentrate on the characteristics of the 100 most massive halos (ranging from $\sim 10^{12} \,{\rm M}_{\odot}$ to $\sim 1.5 \times 10^{14} \, {\rm M}_\odot$) formed at $z=0$. The different aspects we will check in each simulation will be the following:
\begin{itemize}
    \item Visual inspection of the simulation results at $z=0$ (Fig.~\ref{fig:quadri}).
    \item Comparisons of the initial and final power spectra (Fig.~\ref{fig:PS}).
    \item Merger trees and mass accretion histories of the halos formed at $z=0$ (Figs.~\ref{fig:merging} and \ref{fig:example_tree}).
    \item Halo mass function at $z=0$ (Fig.~\ref{fig:HMF}).
    \item Density profiles of the halos and of their environment at $z=0$ (Fig.~\ref{fig:Halo_Cumul}).
    \item Phase-space distribution of subhalos at $z=0$ (Subsection \ref{sec:Subhalo} and Fig.~\ref{fig:satellites}).
\end{itemize}

\section{Results}
\label{sec:resul}

\begin{figure}[ht]
        \centering
        \includegraphics[width=0.5\textwidth]{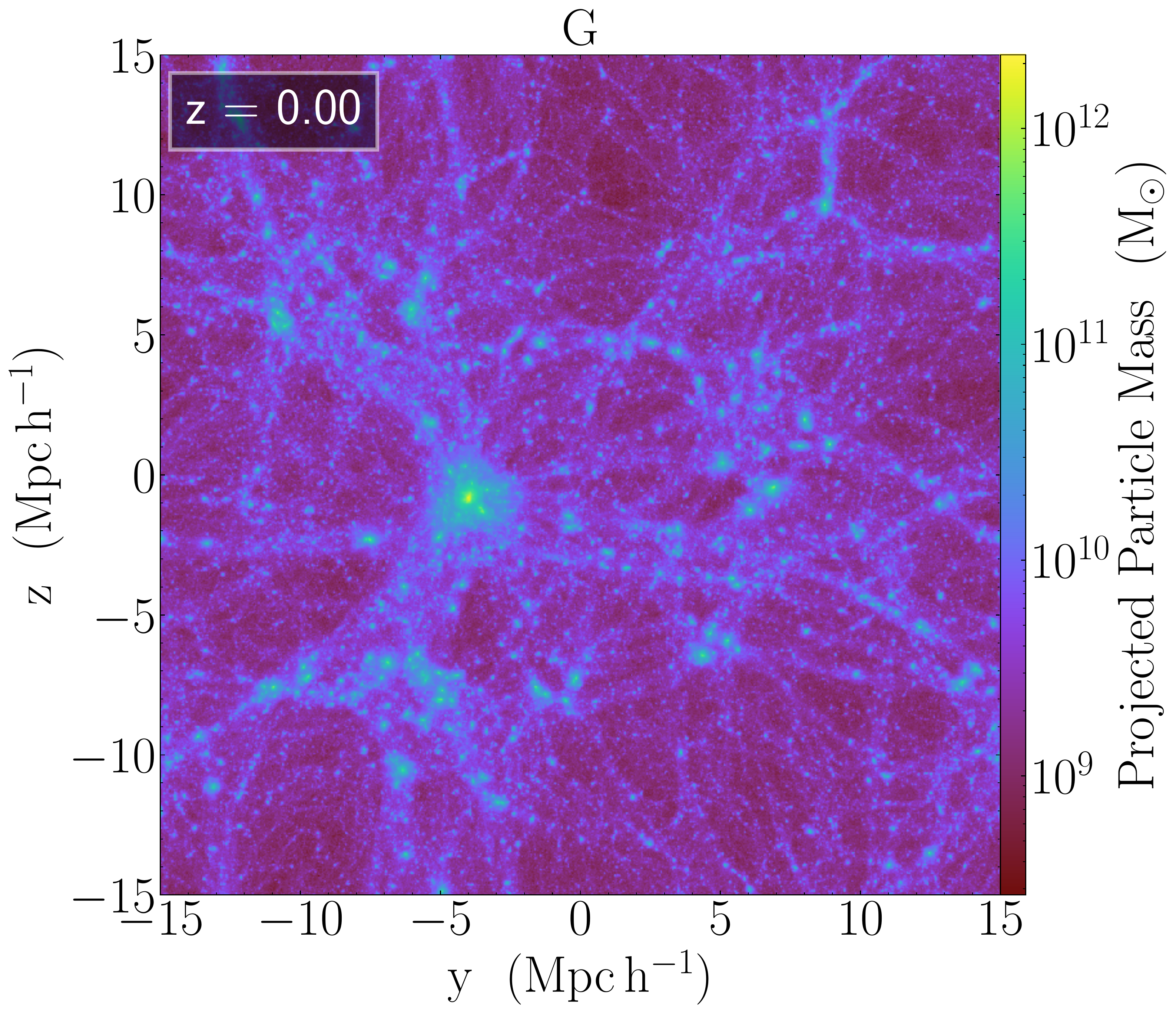}
        \includegraphics[width=0.9\textwidth]{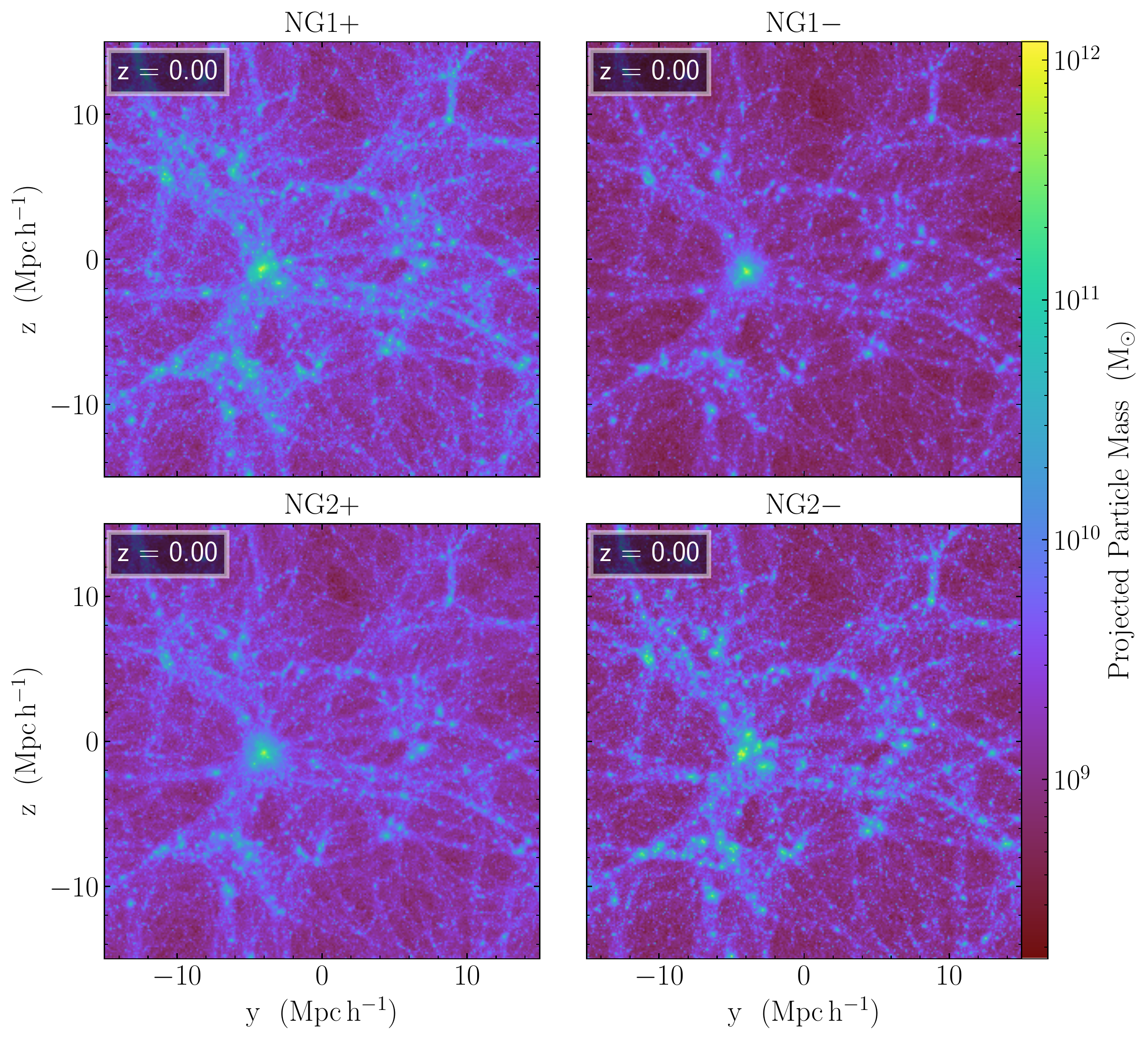}
        \caption{\label{fig:quadri} Visualization of our final configurations at redshift $z=0$. The Gaussian case is shown on top. The two models with skewness are shown in the middle, and the two models with kurtosis at the bottom.}
\end{figure}

We have run our simulations down to $z=0$ and unless stated otherwise, we present all our results at this redshift. We leave it to further studies to study the evolution of the halo mass function and density profiles with redshift. We have identified the halos using the SUBFIND algorithm \cite{Springel:2000qu} included in the public version of \texttt{Gadget 4}. We computed the mass of each halo as $M_{200}$: the mass included in a sphere which corresponds to a mean density of 200 times the critical density of the Universe $\rho_c$. 

In the rest of this section, our main halo sample consists of the 100 most massive halos: their masses range from $\sim 10^{12} \,{\rm M}_{\odot}$ to $\sim 1.5 \times 10^{14} \, {\rm M}_\odot$ in all simulations. More precisely, the Gaussian model G has a mass range of $1.2 \times 10^{12} \,{\rm M}_{\odot}$ to $1.5 \times 10^{14} \, {\rm M}_\odot$, NG1+ has a mass range of $1.3 \times 10^{12} \,{\rm M}_{\odot}$ to $1.4 \times 10^{14} \, {\rm M}_\odot$, NG1- of $1.2 \times 10^{12} \,{\rm M}_{\odot}$ to $1.6 \times 10^{14} \, {\rm M}_\odot$, NG2+ of $1.1 \times 10^{12} \,{\rm M}_{\odot}$ to $1.6 \times 10^{14} \, {\rm M}_\odot$, and NG2- of $1.5 \times 10^{12} \,{\rm M}_{\odot}$ to $1.2 \times 10^{14} \, {\rm M}_\odot$. To test the robustness of our results presented in section \ref{sec:Subhalo}, we have fixed the mass range of our sample of massive halos to be the one of the Gaussian model, and changed the number of halos present in the non-Gaussian samples. All our results presented using our massive halos samples are unchanged. Such a choice of halo sample allows us to explore the inner part of the halos as they are composed of at least 40,000 particles. To ensure that only numerically converged (sub)halos are included in our study, we considered only those halos possessing at least 40 particles, in line with Ref.~\cite{Fakhouri10}.

In Fig.~\ref{fig:quadri}, we first display a visualization of the different models studied in this work. As expected, favouring the overdensities at the initial redshift of the simulation (NG1-) leads to more clumped structure and more empty regions in between. Conversely a simulation favouring underdense regions (NG1+) leads to a final state which is smoother and with in-between regions filled with matter (and, visually, full of substructures). In the case of the kurtosis, one can note that the bimodal NG2- case is visually close to the NG1+ case, but with stronger contrasts related to the bimodal distribution of overdensities. On the other hand, the NG2+ case is visually the closest to the Gaussian case, although with slightly more contrast. 

\subsection{Power spectrum}
\label{sec:PS}
\begin{figure}
        \centering
        \includegraphics[width=1.0\textwidth]{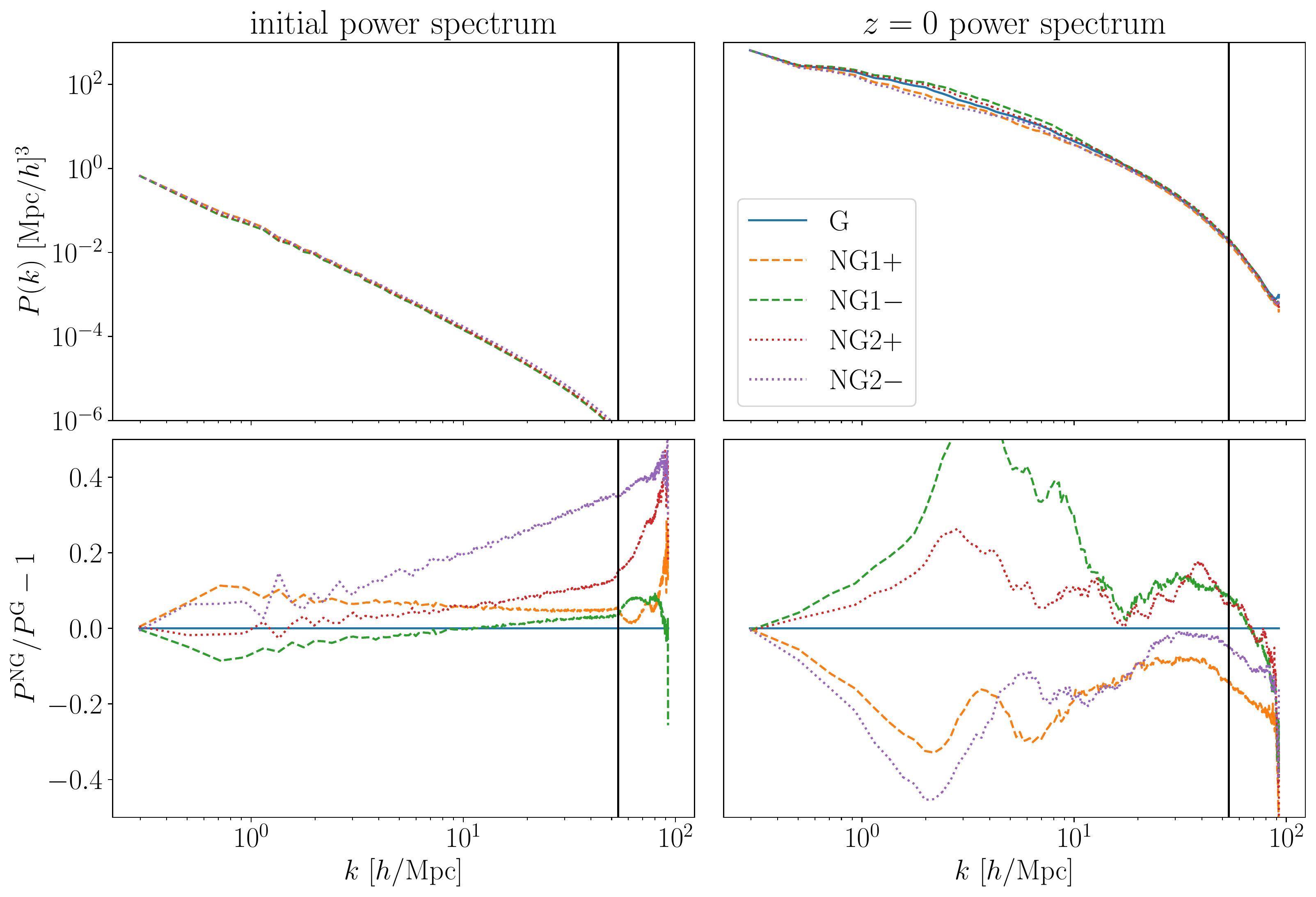}
        \caption{\label{fig:PS} Top panels: Power spectra at initial redshift $z=50$ and final redshift $z=0$ for the different simulations studied in this paper. Bottom panels: relative difference with the Gaussian $\Lambda$CDM power spectrum at $z=50$ and $z=0$, respectively. The vertical line represents the  Nyquist frequency. At $z=0$ NG1- and NG2+ have more power at all intermediate scales, while NG1+ and NG2- do the opposite. This behaviour was not present in the initial power spectrum as one can see from the left panel.}
\end{figure}

Using the library \texttt{Pylians}\footnote{\url{https://github.com/franciscovillaescusa/Pylians}}, we measured the initial and final power spectra of the simulations considered in this paper and compared them with the Gaussian $\Lambda$CDM power spectrum at redshift 0. We present our measurements in Fig.~\ref{fig:PS}. 

Apart from the NG2- case, the initial power spectra are not very different from the Gaussian $\Lambda$CDM case, and they all change over the course of the simulation through the non-linear development of structures. At $z=0$, either the models have systematically less power than $\Lambda$CDM, as for NG1+ or NG2-, or they have systematically more as for NG1- or NG2+. The bump-like feature at $k \sim 3$ $h$/Mpc$\approx k_{\rm NL}$ could potentially lead to interesting observational constraints on $f_{\rm NL}$ independent of the scale dependent bias or the bispectrum. This effect confirms what was seen visually on Fig.~\ref{fig:quadri}: for instance, favouring the overdensities at the initial redshift of the simulation in NG1- leads to more clumped structure and more empty regions in-between, which naturally translates into higher power at $z=0$.  This feature had already been identified\footnote{Annalisa Pillepich, private communication.}
 in earlier numerical work on PNG. This interesting feature of non-Gaussianities may be particularly relevant in the context of the idea to solve the $S_8$ tension with non-linear effects \cite{Amon:2022azi}. In that case, the models NG1+ (with negative $f_{\rm NL}$) and NG2- are the most interesting to investigate. We leave a thorough exploration of this possibility for future work.

\subsection{Merging history}
\label{sec:mergingH}
For each model under study, we considered 29 snapshots between $z=32$ and $z=0$ and constructed the merging history for our halo sample using the library \texttt{ytree} \cite{ytree}. In order to illustrate the scenario depicted in Section \ref{sec:LSSdevelopement}, we compute two quantities:  the `age' of each halo and the merger count. We gather them for each model in Table \ref{table:tree}. To ensure that local correlations do not bias our analysis, we calculated the uncertainties using subbox-based jackknife resampling with 8 boxes. We checked that the standard error on the mean gives comparable estimates.

\begin{figure}
        \centering
        \includegraphics[width=\textwidth]{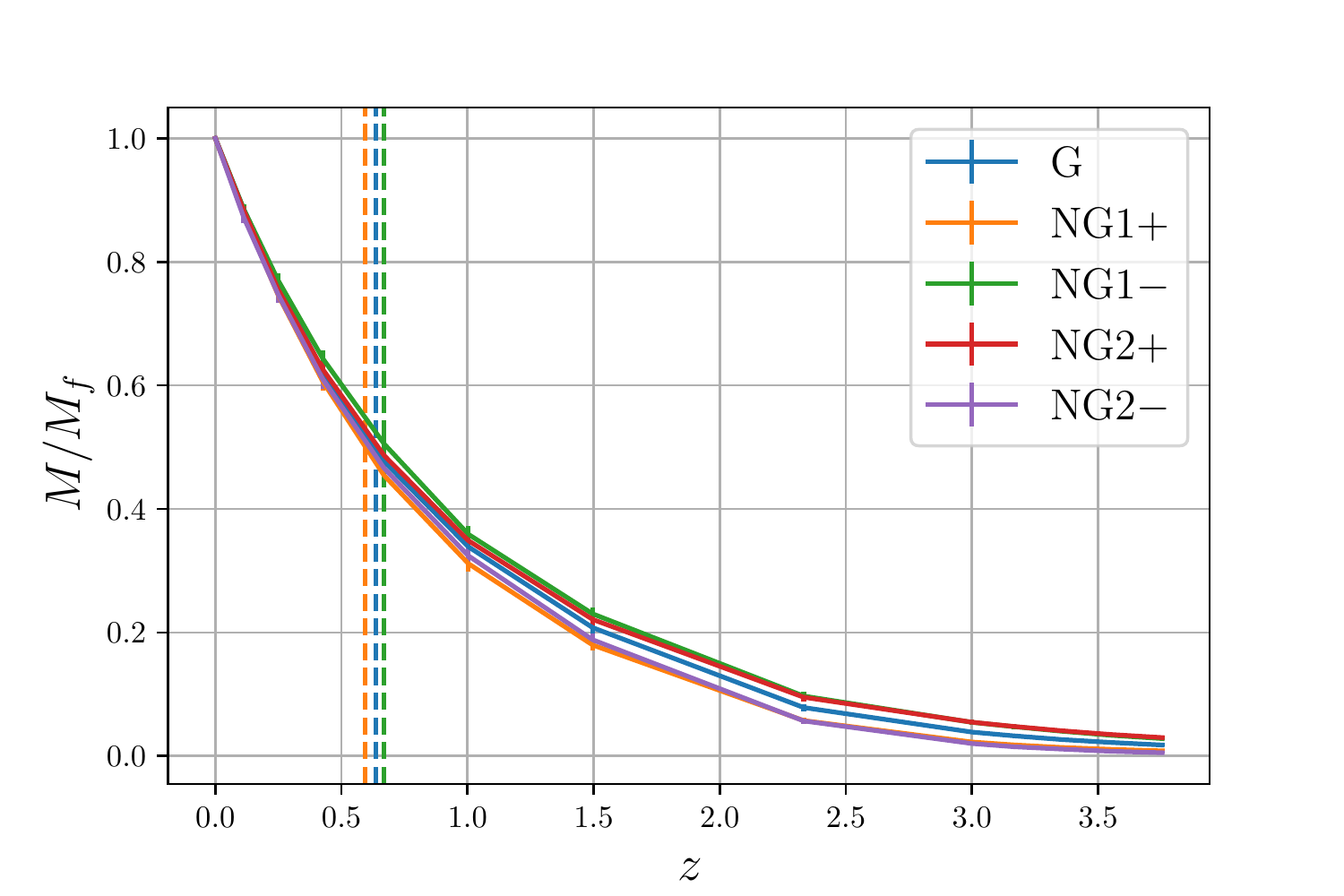}
        \caption{\label{fig:merging} Average ratio of halo mass $M(z)$ over final mass for our halo sample. The vertical lines correspond to the values of $z_{1/2}$ displayed in Table \ref{table:tree} for the Gaussian (G) and skewed (NG1- and NG1+) models. The uncertainties are estimated using the standard error on the mean.}
\end{figure}

We define the `age' of a halo as its half-mass redshift, $z_{1/2}$, the redshift at which the halo weighed half its final mass at $z=0$. Although the difference is not enormous, we do find, as expected, that halos with masses between $\sim 10^{12}$ and $\sim 1.5 \times 10^{14}$ solar masses assemble faster in the NG1- model. In contrast, it takes more time for halos to assemble in the NG1+ case. The complete mass accretion history is displayed in Fig.~\ref{fig:merging} for all simulations.

The merger count MC is defined as the average over our halo sample of the number of mergers between $z=3$ and $z=0$ with halos that featured more than 1$\%$ of the mass of the halo at the time of merging. NG1+ and NG2- have, in this sense, a more violent merging history than the Gaussian one, but all other non-Gaussian models seem to harbour a quieter merging history after $z=3$. This trend follows an initial faster and more violent assembly phase for NG1- and NG2+. 
Indeed, the Gaussian model features on average 0.1 significant mergers at $z>4$, NG1- and NG2+ have typically 0.3 while NG1+ and NG2- have none.

\begin{figure}
        \centering
        \includegraphics[width=\textwidth]{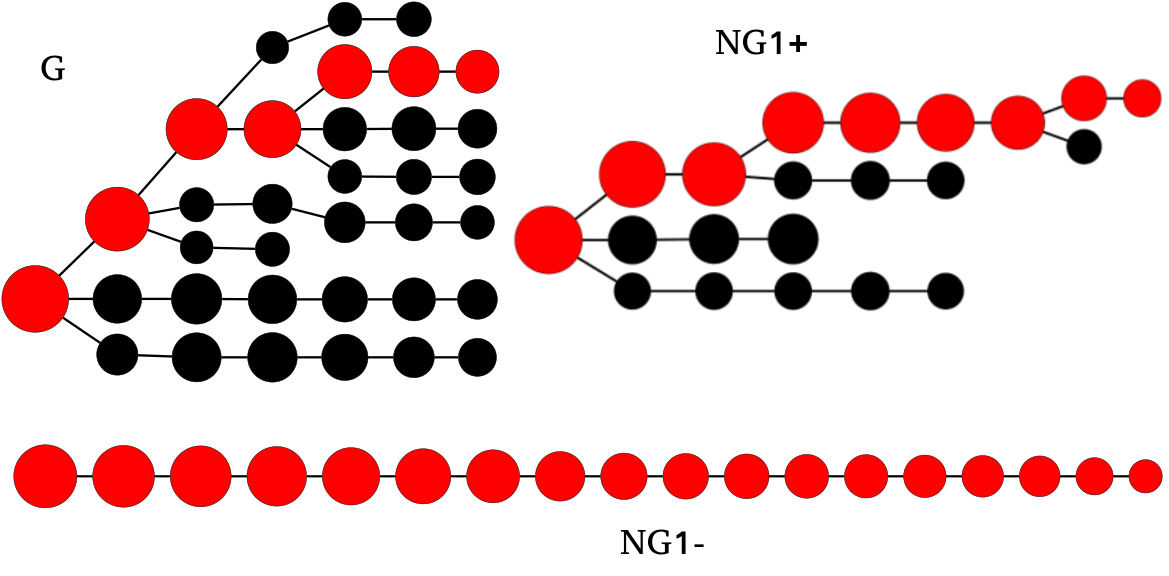}
        \caption{\label{fig:example_tree} The merger trees for the most massive halo of the simulation in the Gaussian case ($M_h= 1.5 \times 10^{14} \,M_{\odot}$) and in the non-Gaussian cases NG1+ ($M_h= 1.4 \times 10^{14} \,M_{\odot}$) and NG1- ($M_h= 1.6 \times 10^{14} \,M_{\odot}$). The radius of the dots are proportional to the logarithm of the mass of the halos, time goes from right to left and we only displayed mergers contributing to at least $1 \text{\textperthousand}$ of the final halo's mass, slightly different than the definition of significant mergers in Table~\ref{table:tree}. As discussed in the main text, in NG1-, the formation of massive halos occurs in a calmer way than in the Gaussian case. Note however that the differences are much more pronounced at such high masses than on average.}
\end{figure}

We display in Fig.~\ref{fig:example_tree} the merging history for the most massive halo we identified in the simulation for the Gaussian model, as well as in the NG1+ and NG1- models. This halo can be found in Fig.~\ref{fig:quadri}, at coordinates (y,z)$\approx$(-4, -1)~Mpc/$h$. Note however that the effect is much more pronounced as such high masses than on average.

\begin{table*}
\begin{center}
\caption{Two quantities describing the merging history: the redshift $z_{1/2}$ at which the halos reach half their $z=0$ mass and the merger count MC which is the sum of all the significant mergers each halo experienced on average between redshift 3 and 0. A significant merger is defined as a merger with mass ratio $>1$\% at the time of merging. The uncertainties are calculated using the jackknife resampling method.} 

 \begin{tabular}{ || l || c | c | c | c |  c || }
\hline\hline  
Simulation & G &  NG1+ &  NG1- & NG2+ & NG2- \\
\hline\hline 
$z_{1/2}$  & 0.64 $\pm 0.01$ & 0.59 $\pm 0.01$ & 0.67 $\pm 0.02$ & 0.64 $\pm 0.01$ & 0.60 $\pm 0.01$ \\
\hline
MC  & 3.5 $\pm 0.1$ & 3.5 $\pm 0.2$ & 3.3 $\pm 0.2$ & 2.8 $\pm 0.2$ & 4.8 $\pm 0.2$ \\
\hline\hline 
\end{tabular}
\label{table:tree}
\vspace{-5mm}
\end{center}
\end{table*}

More thorough studies along the lines of Ref.~\cite{Fakhouri:2008cn,Fakhouri10} are required to further investigate the intriguing merger history formation with non-Gaussian initial conditions. The hints that these merger histories are typically quieter could have important consequences on the formation of bulgeless disks and on bar formation in future hydrodynamical simulations.

\subsection{Halo mass function}
\label{sec:HMF}
\begin{figure}
        \centering
        \includegraphics[width=\textwidth]{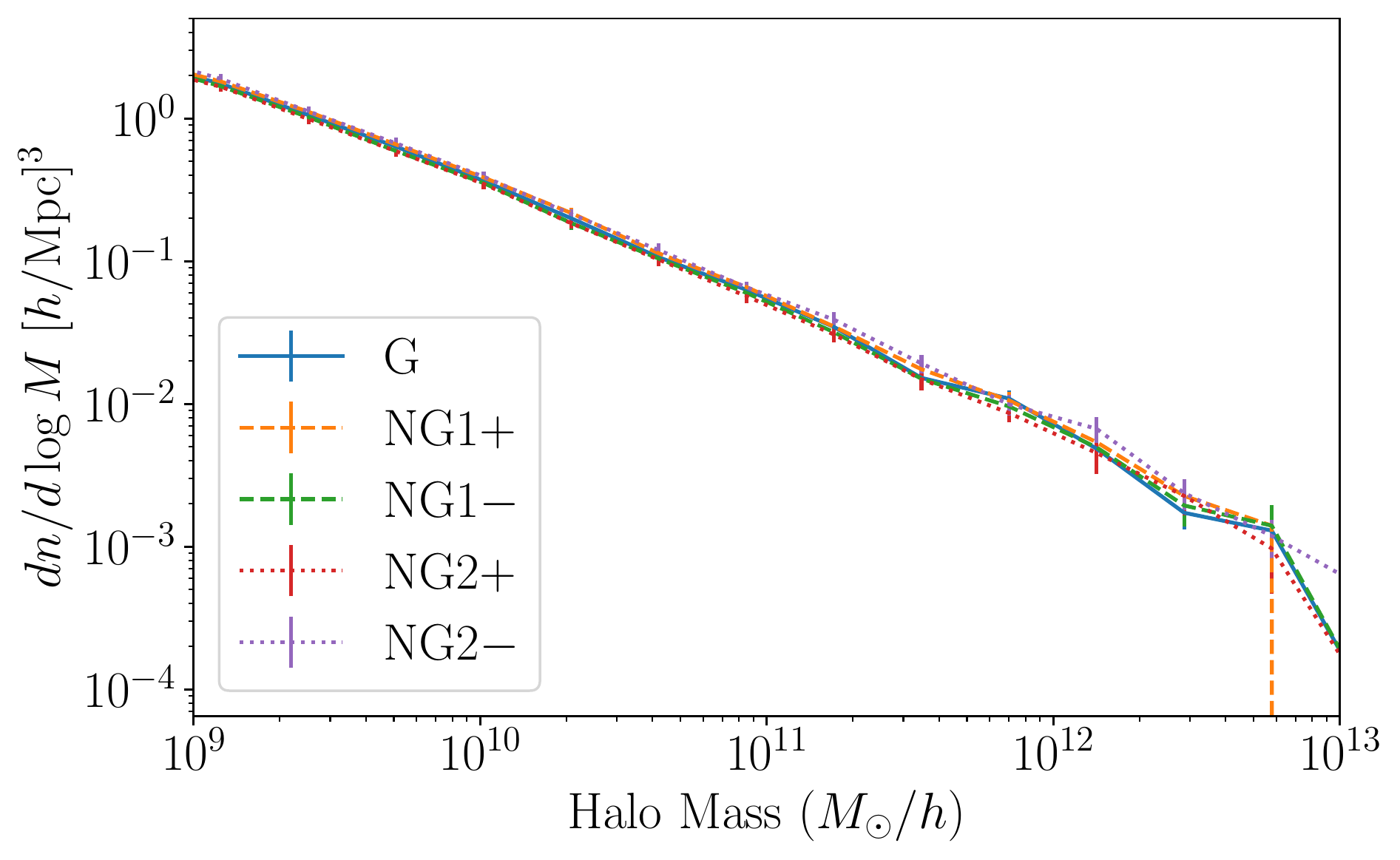}
        \caption{\label{fig:HMF} Halo mass functions for the different models considered. Trends found in previous studies \cite{Dalal:2007cu,Pillepich:2008ka} are recovered, although not in a statistically significant fashion.}
\end{figure}

Extensions of the Press-Schlechter formalism were already put forward to predict the correction to the high mass tail of the Halo mass function (HMF) in a non-Gaussian Universe, see eg.~Ref.~\cite{LoVerde:2007ri}. Those estimates were also backed with N-body simulations, for instance Refs.~\cite{Dalal:2007cu,Pillepich:2008ka} proposed fitting formulae for the HMF in the presence of PNG. In Fig.~\ref{fig:HMF}, we present our ratios of the HMF of the non-Gaussian models studied here with respect to their Gaussian counterpart. The error bars were determined using, as before, subbox-based jackknife resampling.

As anticipated from the visualization (Fig.~\ref{fig:quadri}), NG1+ produces regions between massive halos that are full of substructures, hence more small mass halos than in $\Lambda$CDM. This is also the case for NG2-. On the other hand, NG1- tends to overproduce high mass halos ($>10^{12}{\rm M}_\odot$) and underproduce lower mass halos. This should decrease the typical number of satellites, but also, as anticipated from the visualization, make the typical environment underdense in terms of substructures. Finally, NG2+ overproduces high mass halos too ($>10^{12}{\rm M}_\odot$), and underproduces halos with lower masses, albeit approaching the Gaussian case at masses below $10^{9}{\rm M}_\odot$. Note that these results are however not statistically significant, because we traded good large-box statistics for high resolution. The trends found are however as expected given previous studies with better statistics.

It will be particularly useful, especially for observational tests of the models with luminosity functions at high redshift, to compute the HMF as a function of redshift, which we postpone to future work, since the present paper mostly concentrates on the structure of halos at $z=0$. This also requires to move to a scale-dependent $f_{\rm NL}$ to be able to go to larger boxes and have more statistics as well as extreme events.
Ref.~\cite{Biagetti:2022ode} has very recently already studied several aspects of this question in the context of the possibility of unexpectedly massive galaxies at high redshift observed with JWST \cite{Labbe, MBK2}.
\subsection{The inner structure of halos and their environment}

\subsubsection{Density profiles of the halos and their surroundings}
The effect of PNG on the halo density profiles has been studied in the past \cite{Avila-Reese:2003cjm,Smith:2010fh,MoradinezhadDizgah:2013rkr}, where the effect of positively/negatively skewed non-Gaussianities in terms of density contrast resulted in more/less concentrated halos. As discussed in the introduction several small scale problems of cosmological structure formation might possibly be alleviated if halos were formed in less dense environments, where they would in particular perhaps allow disks to grow more quietly without bulge formation through their quieter merging history, while at the same time not preventing bar formation. It is already known that bars form naturally in isolated idealized galaxy simulations, and are inhibited only in a cosmological context.  Hence, having a more isolated environment would allow cosmological simulations to resemble more closely the idealized simulations where bar formation is not a problem.

\begin{figure}
        \centering
        \includegraphics[width=\textwidth]{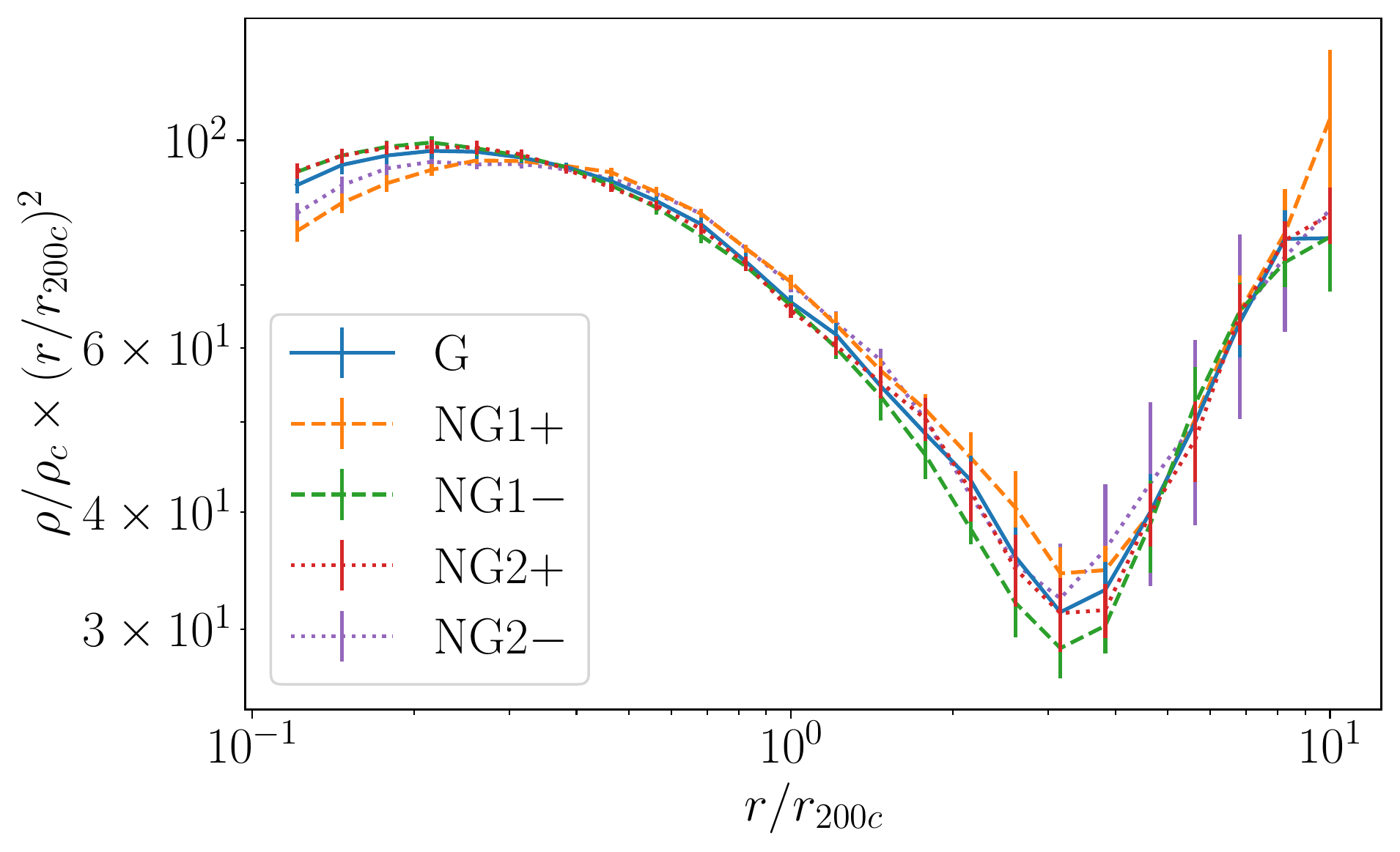}
        \caption{\label{fig:Halo_Cumul} Stacked density of dark matter in concentric spheres centred on the 100 more massive halos all the way to $10 \times r_{\rm 200c}$. For visibility, we have multiplied the density by $r^2/r_{200c}^2$. 
        }
\end{figure}

To gauge the potential impact of non-Gaussianity on these properties of the inner regions of galaxies, we chose to examine the density profiles out to a large distance of $10 \times r_{\rm 200c}$, where $r_{\rm 200c}$ is the scale at which the inner mean density of the halo is 200 times the critical density of the universe. We stacked the density profiles of the 100 most massive halos identified in the simulation, and we present their average in Fig.~\ref{fig:Halo_Cumul}. The uncertainties were calculated using sub-box based jackknife resampling, as before. We have checked by separating the 100 halos in 10 mass bins that this choice of mass does not impact the results presented in this section.

Concerning the inner structure of halos, it appears that the NG1- and NG2+ model are slightly more concentrated than the Gaussian case, while NG1+ and NG2- are less concentrated. The trend is then reversed for the immediate surroundings of the halos, denser for NG1+ and NG2- and less dense for NG1- and NG2+: this is most clear in the NG1- model, where the environment between 0.5 and 4 virial radii of the halos is underdense compared to $\Lambda$CDM. A similar tendency was also pointed out e.g.~in Ref.~\cite{Dalal:2008zd}, see also Ref.~\cite{Lazeyras:2022koc}. This trend, coupled with the slightly quieter merging history found for the NG1- halos, makes this model particularly appealing for further investigations of disk and bar formation in future hydrodynamical simulations. However, whilst the trend is visually clear and recovers similar results from previous contributions \cite{Avila-Reese:2003cjm,Dalal:2008zd}, it is not statistically significant in our small-box simulations.

\subsubsection{Kinematic coherence of satellite halos}
\label{sec:Subhalo}

As briefly mentioned in the introduction, the observed phase-space distribution of satellite galaxies within their host galaxy's halo, both in the Local Group and beyond \citep[e.g.,][]{Pawlowski1, Ibata:2014csa, Muller}, presents one of the most pressing challenges to $\Lambda$CDM on sub-Mpc scales, largely independent of baryonic physics and even to various modifications of the nature of dark matter, the latter being more prone to modifying the internal structure of subhalos or their abundance than their actual phase-space distribution. 

This problem has recently been reviewed in Ref.~\cite{Pawlowski2}: see references therein for a comprehensive overview of the problem. In the Milky Way, the peculiar distribution of satellites was already noticed in the seventies \cite{LyndenBell}, and is nowadays known as the `Vast Polar Structure' \citep[VPOS,][]{Pawlowski3}, a rotating plane of satellite galaxies, which also contains several globular clusters. The flattening of this structure can be quantified by, e.g., the minor-to-major axis ratio $c/a$ of the 11 brightest satellite galaxies, whose value is $c/a=0.182$ \citep{Pawlowski3}, hence a significantly flattened structure, not typical in simulated $\Lambda$CDM halos \cite{Wang:2012bt}. What is more, Gaia proper motions have shown that 50\% to 75\% of the satellites within the VPOS are orbiting within this structure \citep{LiH}. This is not the case of the typical distribution of subhalos in $\Lambda$CDM simulations. For the Milky Way, such values have been claimed to be in accordance with recent hydrodynamical simulations when taking into account the high central concentration of the Milky Way system of satellites, combined with the closeness of its two most distant members \cite{Sawala:2022xom}, but it is worth noting that satellite planes have also been found around M31 \citep{IbataM31} and Cen A \citep{Muller}, and hints of kinematical coherence have been found in SDSS data for systems containing at least two observed satellites with line-of-sight velocities \citep{Ibata:2014csa}.

In this subsection, we first confirm that the $c/a$ ratio is close to values obtained in previous simulations \cite{Wang:2012bt} and that there is no sign of kinematic coherence in our Gaussian $\Lambda$CDM simulation. We will then check whether non-Gaussian initial conditions can improve the situation, and which type of non-Gaussianities would be preferred in that sense. While previous diagnostics of our non-Gaussian setup were studied before (at least with larger box sizes), the study of the kinematic coherence of halos formed with non-Gaussian initial conditions is performed for the first time in this work. 

We perform the following analysis with our 100 halos sample: if they exist, we consider their 11 most massive subhalos as detected by SUBFIND. We then compute the inertia tensor:
\be
I_{ij}=\sum_{n=1}^N x_{n,i} x_{n,j},
\ee
where the $\mathbf{x_{n}}$ are the three-dimensional coordinates of the $N=11$ subhalos relative to the central halo.
The ordered square roots of the eigenvalues of the inertia tensor are labelled $a$, $b$ and $c$, and the quantity we will compute for our halo sample is the mean over the 100 most massive halos of the ratio $c/a$.  Table \ref{table:coplan} displays our values of $c/a$ for each model. All models have similar flattenings within errors, apart from the NG2- model which is less flattened than the Gaussian case. The slightly more flattened distribution in the NG1- model is not significant. The values are close to those obtained in previous Gaussian simulations \cite{Wang:2012bt}.

\begin{table*}
\begin{center}
\caption{Three quantities describing the flattening and kinematic coherence of subhalos around the 100 most massive host halos of our simulation. The average minor-to-major axis ratio $c/a$ of the 11 most massive subhalos around each host. The uncertainties were calculated using the same jackknife resampling method as in Section \ref{sec:mergingH} and were checked to be similar to the standard error on the mean. The ratio AC/C of the number of anti-correlated over correlated line-of-sight velocities for two diametrically opposed most massive subhalos and three different views of the host halos is also listed. The latter quantity is computed for two tolerance angles $\alpha=10^\circ$ and $\alpha=50^\circ$ ; the error bars were calculated using integrals of the beta distribution  (see text for details).} 

 \begin{tabular}{ || l || c | c | c | c |  c || }
\hline\hline  
Simulation & G &  NG1+ &  NG1- & NG2+ & NG2- \\
\hline\hline 
mean $c/a$  & 0.33 $\pm$ 0.01 & 0.34 $\pm$ 0.01 & 0.32 $\pm$ 0.02 & 0.31 $\pm$ 0.01 & 0.37 $\pm$ 0.02 \\
\hline
AC/C,  $\alpha=10^\circ$ & $1.1 \pm 0.5$ & $1.0 \pm 0.8$ & $1.3 \pm 0.6$ & $0.9 \pm 0.6$ & $1.0 \pm 0.8$ \\
\hline
AC/C,  $\alpha=50^\circ$  & $0.9 \pm 0.2$ & $0.8 \pm 0.2$ & $1.5 \pm 0.3$ & $0.9 \pm 0.2$ & $1.0 \pm 0.2$ \\
\hline \hline
\end{tabular}
\label{table:coplan}
\vspace{-5mm}
\end{center}
\end{table*}

\begin{figure*}
        \centering
        \begin{tabular}[t]{cc}
        \includegraphics[width=0.5\textwidth]{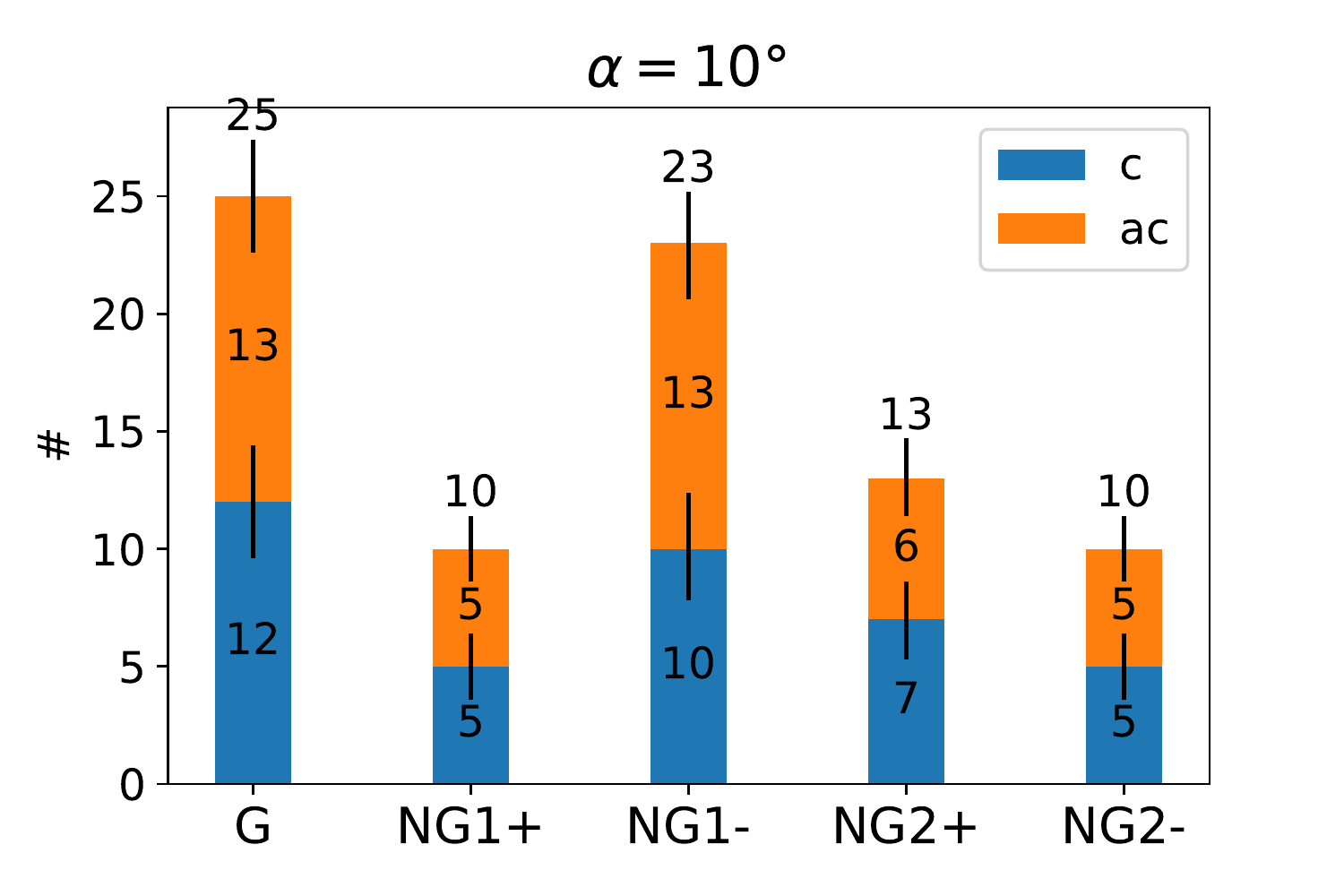} &
        \includegraphics[width=0.5\textwidth]{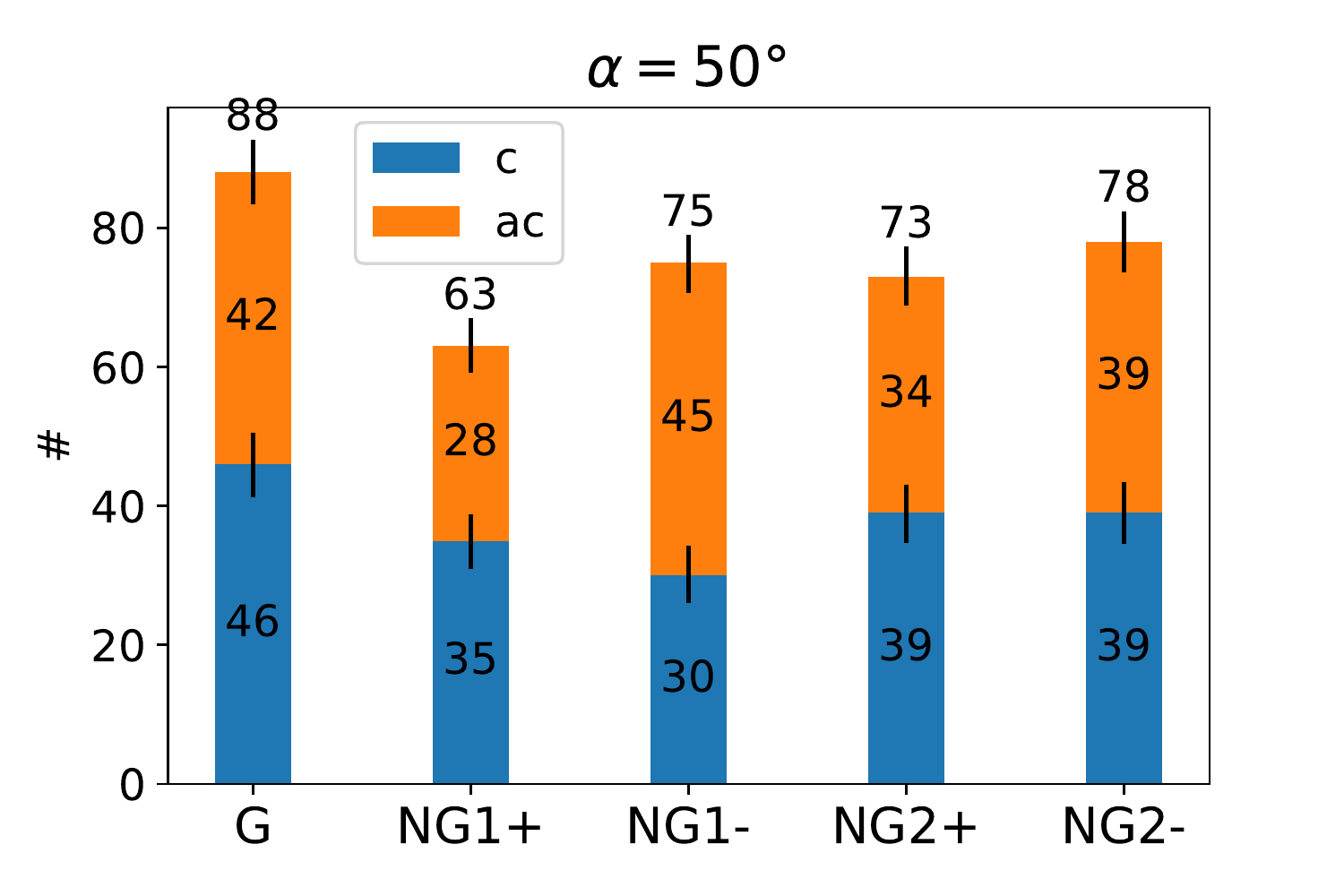}
        \end{tabular}
        \caption{Histogram over 3 views of the number of anti-correlated and correlated line-of-sight velocities for two diametrically opposed most massive subhalos (within a tolerance angle $\alpha$, see text for details). For $\alpha=50^{\circ}$ and NG1-, there is a trend of kinematic coherence (with a larger number of anti-correlated systems over correlated ones), robust to the error-bars represented with vertical black lines.} \label{fig:satellites}
\end{figure*}

Next, we investigate the other main open question for satellite halos: their kinematic coherence. After finding the best-fitting plane to the 11 most massive subhalos, no clear sign of kinematic coherence along this plane was found. However, concentrating on the two most massive subhalos, similar to the test performed on SDSS data \cite{Ibata:2014csa}, a clear trend does emerge. For this test, we first identify the most massive subhalo around each of our 100 host halos. We then choose three different views of those host halos, and we draw within the `sky plane' a line connecting the most massive subhalo to the host halo. We then check if the second most massive subhalo lies on the opposite side within a tolerance angle $\alpha$ defined as the maximum allowed angle between the line connecting the most massive subhalo and the host and the line connecting the second most massive subhalo and the host. \citep[see Fig.~2 of ][]{Ibata:2014csa}. If the second most massive subhalo lies within the tolerance angle $\alpha$, we check whether it shares the same sign of the line-of-sight velocity (correlated = C) or the opposite sign (anti-correlated = AC) with the most massive subhalo. The histograms of C and AC are shown in Fig.~\ref{fig:satellites}, and the values of the ratios AC/C are reported in Table \ref{table:coplan} for all our simulations and for two tolerance angles: $10^\circ$ and $50^\circ$. Ref.~\cite{Ibata:2014csa} reported a value AC/C=2.4 at $15^\circ$ from SDSS data. Our uncertainties are calculated at 1$\sigma$ using binomial confidence intervals computed with integrals of the Beta distribution \cite{Cameron:2010bh}. As expected, this ratio is very close to 1 in the Gaussian $\Lambda$CDM case. It is also the case in most other non-Gaussian cases, but the only exception is NG1- which exhibits a  trend of kinematic coherence at nearly 2$\sigma$ that makes it the most promising non-Gaussian model to alleviate the kinematic coherence of satellites tension with $\Lambda$CDM. With the statistics available to us in the present simulations, it is hard to draw definitive conclusions, but this result motivates the future design of larger-box simulations with scale-dependent non-Gaussian initial conditions, followed by hydrodynamical zoom-in simulations to investigate how a large statistical sample of galaxy satellites will behave in the NG1- case. 

\section{Conclusions}
\label{sec:concl}

We began this contribution by summarising current observational constraints on non-Gaussianities, and underlining the fact that on scales of the order of $\sim 10$ Mpc, the stringent observational constraints on large-scale non-Gaussianity do not apply. Motivated by these considerations, we investigated with gravity-only collisionless simulations the possible consequences of a change to the initial conditions on the formation of typical galaxy-mass halos. This was achieved by numerically following structure formation in a box of 30 Mpc/$h$.

We explored four different types of non-Gaussianities: two models with skewness, and two models with kurtosis. The model with a positive skewness in terms of the gravitational potential, dubbed NG1+, translates into a large tail for negative $\delta$ (hence a negative $f_{\rm NL}$), and the mode of the distribution peaks at a small overdensity. The case with negative skewness in terms of the potential, which we dub NG1-, translates into a large tail towards large positive overdensities (positive $f_{\rm NL}$) and the mode of the distribution peaking at a slight underdensity. The model with kurtosis with a single bump centred on $0$ is dubbed NG2+, while the bimodal model with kurtosis, characterised by two bumps around $0$, is dubbed NG2-. If one wishes to increase the amount of overdensities with $\delta \sim 0.5$ and above, with respect to the Gaussian case, the two relevant scenarios are NG1- and NG2+.

We compared the initial and final power spectra and found that, at $z=0$ NG1- and NG2+ have more power than the Gaussian case, while NG1+ and NG2- display the opposite behaviour. This feature may be particularly interesting in view of the $S_8$ tension and the phenomenological parametrization of Ref.~\cite{Amon:2022azi}. In terms of the halo mass function, NG1- and NG2+ tend to overproduce high mass halos ($>10^{12}{\rm M}_\odot$) and underproduce lower mass halos. This decreases the typical number of satellites, but also makes the typical environment underdense in terms of substructures. The opposite happens for NG1+ and NG2-. We mostly focused on the 100 most massive halos present at redshift $z=0$ in our simulations, with a mass range from $\sim 10^{12} \,M_{\odot}$ to $\sim 1.5 \times 10^{14}$. We showed that, in terms of the inner density profiles, NG1- and NG2+ produce slightly more concentrated halos than in the Gaussian case, while the opposite happens for NG1+ and NG2-. The density of the nearby environment (0.5 to 4 virial radii) is on the other hand lower than in the Gaussian case for NG1- and NG2+, with a most clear effect for the NG1- case. This is particularly appealing in view of alleviating some current challenges of galaxy formation such as bar formation in a cosmological context, which is known to pose no problem in an idealized isolated environment. A quieter merging history might also help alleviating the `hot orbits' problem of current simulations. We showed that the NG1+ and NG2- models typically harbour halos with a more violent merging history than in the Gaussian case, but all other non-Gaussian models seem to harbour a quieter merging history at $z<3$. Although the difference is not enormous, we also find that halos with masses between $\sim 10^{12}$ and $\sim 1.5 \times 10^{14}$ solar masses assemble faster in the NG1- model, whilst, on the contrary, it takes more time for halos to assemble in the NG1+ and NG2- cases. A quiet merging history coupled with a fast mass assembly makes the NG1- model particularly appealing to potentially solve both the `hot orbits' problem and the missing bar problem. Moreover, we have shown that this NG1- model shows possible signs of kinematic coherence of subhalos around their host, thereby possibly alleviating the satellite phase-space correlation problem of $\Lambda$CDM. However, our results for this model only alleviate the phase-space correlation tension without solving it completely, and will certainly need larger size simulations coupled with hydrodynamics to explore whether there is a chance of solving it with PNG alone.

Having characterised all those effects of PNG on galactic scales, the next pressing question we want to answer is how those effects translate when one considers state-of-the art zoom hydrodynamical simulations of Milky Way sized galaxies in this context. The potential of the minimal changes in initial conditions that we performed here could indeed translate into drastically different bars, bulges and stellar halos in hydrodynamical simulations compared to the Gaussian case. 

Another obvious improvement will be to devise scale-dependent non-Gaussian initial conditions instead of the effective local non-Gaussian templates used here. This will allow us to probe the transition from a Gaussian Universe on large scales to a non-Gaussian one on small scales, and to run larger box simulations vastly improving our current statistics. Voids in this cosmology would be also a very interesting probe to investigate as they carry more linear information or could be a cleaner probe of the environmental structure of the halos.  JWST preliminary results may have shown bright massive galaxies at high redshift that are very difficult to account for within $\Lambda$CDM. Scale dependent PNG could also be preferred over $\Lambda$CDM in such a context \cite{Biagetti:2022ode}.

While a low pass filter or a power-law are natural candidates to improve our initial condition templates, inflationary physics may provide more sound candidates \cite{Dalal11} and reciprocally, our studies on small scales could possibly constrain or motivate certain models. In particular an approach that has been gaining some momentum in recent years is to get rid of the perturbative expansion around a Gaussian PDF and to consider the full non-Gaussian PDF, see eg. Ref.~\cite{Pena:2022sdg} and references therein. 

The minimal changes to the initial conditions considered in this paper are obviously not mutually exclusive with other modifications of the dark sector, but we choose to concentrate here on these to isolate their potential effects. For instance, a minimal modification to our present treatment would be to follow the suggestion of Ref.~\cite{Peebles:2020bph} advocating for a warm dark matter scenario with non-Gaussian initial conditions. While some models involving non-Gaussianity would also need to take into account other radical changes in the dynamics of the dark sector \citep[e.g.,][]{stahlDDM}, we note that many dark matter candidates only require some additional small changes to initial conditions such as power spectra cutoffs or oscillations: this means that the exploration of various dark matter candidates with, e.g., ETHOS type parameterization \cite{Murgia:2017lwo} might require only minimal changes to the existing codes used here, and could therefore easily be combined with non-Gaussianities.

\acknowledgments
BF, RI and CS acknowledge funding from the European Research Council (ERC) under the European Union's Horizon 2020 research and innovation program (grant agreement No.\ 834148). TM and OH acknowledge funding from the European Research Council (ERC) under the European Union's Horizon 2020 research and innovation program (grant agreement No.\ 679145). CS and BF acknowledge early discussions on the topic with Martin Bordeau, Nicolas Mai, Thomas Oliveira and Renaud Vancoellie. CS also appreciates a nice discussion with Sébastien Renaux-Petel in the TUG 2021 workshop. CS acknowledges insightful exchanges with Vladimir Avila-Reese. OH acknowledges discussions on PNG with Annalisa Pillepich, who also pointed out the PNG PS bump to us. This work has made use of the Infinity Cluster hosted by the Institut d'Astrophysique de Paris. The analysis was partially made using the YT python package \cite{Turk:2010ah} and a development of ytree \cite{ytree}, as well as IPython \cite{Perez:2007emg}, Matplotlib \cite{Hunter:2007ouj} and NumPy \cite{vanderWalt:2011bqk}. We acknowledge the anonymous referee for an exceptionally clear and constructive report, which greatly improved the manuscript.
\section*{Authors' Contribution}
This work is based on an original idea put forward by BF. TM implemented in \texttt{MonofonIC} the non-Gaussian initial conditions helped by OH. The simulations presented in this work were performed and analysed by CS and TM, in consultation with BF and OH. CS and BF drafted the manuscript, helped by TM for section \ref{sec:NGIC}. RI and OH improved it heavily by their comments. 

\section*{Carbon Footprint}
Following Ref.~\cite{berthoud} to convert\footnote{Including the global utilisation of the cluster and the pollution due to the electrical source, the conversion factor is 4.7 gCO2e/h core} the number of CPU hours required to obtain the data for this work, we have used 3 tCO2eq.

\bibliographystyle{JHEP.bst}
\bibliography{ref.bib}

\providecommand{\href}[2]{#2}\begingroup\raggedright\begin{thebibliography}{10}

\bibitem{Ostriker}
J.~P. {Ostriker} and P.~J. {Steinhardt}, \emph{{The observational case for a
  low-density Universe with a non-zero cosmological constant}},
  \href{http://dx.doi.org/10.1038/377600a0}{\emph{\nat} {\bf 377} (Oct., 1995)
  600--602}.

\bibitem{Planck:2018vyg}
{\scshape Planck} collaboration, N.~Aghanim et~al., \emph{{Planck 2018 results.
  VI. Cosmological parameters}},
  \href{http://dx.doi.org/10.1051/0004-6361/201833910}{\emph{Astron.
  Astrophys.} {\bf 641} (2020) A6},
  [\href{https://arxiv.org/abs/1807.06209}{{\tt 1807.06209}}].

\bibitem{Abdalla}
E.~{Abdalla}, G.~F. {Abell{\'a}n}, A.~{Aboubrahim}, A.~{Agnello},
  {\"O}.~{Akarsu}, Y.~{Akrami} et~al., \emph{{Cosmology intertwined: A review
  of the particle physics, astrophysics, and cosmology associated with the
  cosmological tensions and anomalies}},
  \href{http://dx.doi.org/10.1016/j.jheap.2022.04.002}{\emph{Journal of High
  Energy Astrophysics} {\bf 34} (June, 2022) 49--211},
  [\href{https://arxiv.org/abs/2203.06142}{{\tt 2203.06142}}].

\bibitem{Famaey}
B.~{Famaey} and S.~S. {McGaugh}, \emph{{Modified Newtonian Dynamics (MOND):
  Observational Phenomenology and Relativistic Extensions}},
  \href{http://dx.doi.org/10.12942/lrr-2012-10}{\emph{Living Reviews in
  Relativity} {\bf 15} (Sept., 2012) 10},
  [\href{https://arxiv.org/abs/1112.3960}{{\tt 1112.3960}}].

\bibitem{Bullock}
J.~S. {Bullock} and M.~{Boylan-Kolchin}, \emph{{Small-Scale Challenges to the
  {\ensuremath{\Lambda}}CDM Paradigm}},
  \href{http://dx.doi.org/10.1146/annurev-astro-091916-055313}{\emph{\araa}
  {\bf 55} (Aug., 2017) 343--387},
  [\href{https://arxiv.org/abs/1707.04256}{{\tt 1707.04256}}].

\bibitem{Peebles2022}
P.~J.~E. {Peebles}, \emph{{Anomalies in Physical Cosmology}}, {\emph{arXiv
  e-prints} (Aug., 2022) arXiv:2208.05018},
  [\href{https://arxiv.org/abs/2208.05018}{{\tt 2208.05018}}].

\bibitem{Auriga}
R.~J.~J. {Grand}, F.~A. {G{\'o}mez}, F.~{Marinacci}, R.~{Pakmor},
  V.~{Springel}, D.~J.~R. {Campbell} et~al., \emph{{The Auriga Project: the
  properties and formation mechanisms of disc galaxies across cosmic time}},
  \href{http://dx.doi.org/10.1093/mnras/stx071}{\emph{\mnras} {\bf 467} (May,
  2017) 179--207}, [\href{https://arxiv.org/abs/1610.01159}{{\tt 1610.01159}}].

\bibitem{Fire}
S.~{Garrison-Kimmel}, P.~F. {Hopkins}, A.~{Wetzel}, K.~{El-Badry}, R.~E.
  {Sanderson}, J.~S. {Bullock} et~al., \emph{{The origin of the diverse
  morphologies and kinematics of Milky Way-mass galaxies in the FIRE-2
  simulations}}, \href{http://dx.doi.org/10.1093/mnras/sty2513}{\emph{\mnras}
  {\bf 481} (Dec., 2018) 4133--4157},
  [\href{https://arxiv.org/abs/1712.03966}{{\tt 1712.03966}}].

\bibitem{Pillepich}
A.~{Pillepich}, V.~{Springel}, D.~{Nelson}, S.~{Genel}, J.~{Naiman},
  R.~{Pakmor} et~al., \emph{{Simulating galaxy formation with the IllustrisTNG
  model}}, \href{http://dx.doi.org/10.1093/mnras/stx2656}{\emph{\mnras} {\bf
  473} (Jan., 2018) 4077--4106}, [\href{https://arxiv.org/abs/1703.02970}{{\tt
  1703.02970}}].

\bibitem{Vogelsberger:2019ynw}
M.~Vogelsberger, F.~Marinacci, P.~Torrey and E.~Puchwein, \emph{{Cosmological
  Simulations of Galaxy Formation}},
  \href{http://dx.doi.org/10.1038/s42254-019-0127-2}{\emph{Nature Rev. Phys.}
  {\bf 2} (2020) 42--66}, [\href{https://arxiv.org/abs/1909.07976}{{\tt
  1909.07976}}].

\bibitem{Dubois}
Y.~{Dubois}, R.~{Beckmann}, F.~{Bournaud}, H.~{Choi}, J.~{Devriendt},
  R.~{Jackson} et~al., \emph{{Introducing the NEWHORIZON simulation: Galaxy
  properties with resolved internal dynamics across cosmic time}},
  \href{http://dx.doi.org/10.1051/0004-6361/202039429}{\emph{\aap} {\bf 651}
  (July, 2021) A109}, [\href{https://arxiv.org/abs/2009.10578}{{\tt
  2009.10578}}].

\bibitem{Angulo:2021kes}
R.~E. Angulo and O.~Hahn, \emph{{Large-scale dark matter simulations}},
  \href{https://arxiv.org/abs/2112.05165}{{\tt 2112.05165}}.

\bibitem{Pawlowski1}
M.~S. {Pawlowski}, \emph{{The planes of satellite galaxies problem, suggested
  solutions, and open questions}},
  \href{http://dx.doi.org/10.1142/S0217732318300045}{\emph{Modern Physics
  Letters A} {\bf 33} (Feb., 2018) 1830004},
  [\href{https://arxiv.org/abs/1802.02579}{{\tt 1802.02579}}].

\bibitem{Ibata:2014csa}
N.~G. Ibata, R.~A. Ibata, B.~Famaey and G.~F. Lewis, \emph{{Velocity
  anti-correlation of diametrically opposed galaxy satellites in the low
  redshift universe}},
  \href{http://dx.doi.org/10.1038/nature13481}{\emph{Nature} {\bf 511} (2014)
  563}, [\href{https://arxiv.org/abs/1407.8178}{{\tt 1407.8178}}].

\bibitem{Muller}
O.~{M{\"u}ller}, M.~S. {Pawlowski}, H.~{Jerjen} and F.~{Lelli}, \emph{{A
  whirling plane of satellite galaxies around Centaurus A challenges cold dark
  matter cosmology}},
  \href{http://dx.doi.org/10.1126/science.aao1858}{\emph{Science} {\bf 359}
  (Feb., 2018) 534--537}, [\href{https://arxiv.org/abs/1802.00081}{{\tt
  1802.00081}}].

\bibitem{Katz91}
N.~{Katz} and J.~E. {Gunn}, \emph{{Dissipational Galaxy Formation. I. Effects
  of Gasdynamics}}, \href{http://dx.doi.org/10.1086/170367}{\emph{\apj} {\bf
  377} (Aug., 1991) 365}.

\bibitem{Navarro91}
J.~F. {Navarro} and W.~{Benz}, \emph{{Dynamics of Cooling Gas in Galactic Dark
  Halos}}, \href{http://dx.doi.org/10.1086/170590}{\emph{\apj} {\bf 380} (Oct.,
  1991) 320}.

\bibitem{DOnghia:2004iex}
E.~D'Onghia and A.~Burkert, \emph{{Bulgeless galaxies and their angular
  momentum problem}}, \href{http://dx.doi.org/10.1086/424444}{\emph{Astrophys.
  J. Lett.} {\bf 612} (2004) L13--L16},
  [\href{https://arxiv.org/abs/astro-ph/0402504}{{\tt astro-ph/0402504}}].

\bibitem{Peebles:2020bph}
P.~J.~E. Peebles, \emph{{Formation of the Large Nearby Galaxies}},
  \href{http://dx.doi.org/10.1093/mnras/staa2649}{\emph{Mon. Not. Roy. Astron.
  Soc.} {\bf 498} (2020) 4386--4395},
  [\href{https://arxiv.org/abs/2005.07588}{{\tt 2005.07588}}].

\bibitem{Reddish}
J.~{Reddish}, K.~{Kraljic}, M.~S. {Petersen}, K.~{Tep}, Y.~{Dubois},
  C.~{Pichon} et~al., \emph{{The NewHorizon simulation - to bar or not to
  bar}}, \href{http://dx.doi.org/10.1093/mnras/stac494}{\emph{\mnras} {\bf 512}
  (May, 2022) 160--185}, [\href{https://arxiv.org/abs/2106.02622}{{\tt
  2106.02622}}].

\bibitem{Roshan}
M.~{Roshan}, N.~{Ghafourian}, T.~{Kashfi}, I.~{Banik}, M.~{Haslbauer},
  V.~{Cuomo} et~al., \emph{{Fast galaxy bars continue to challenge standard
  cosmology}}, \href{http://dx.doi.org/10.1093/mnras/stab2553}{\emph{\mnras}
  {\bf 508} (Nov., 2021) 926--939},
  [\href{https://arxiv.org/abs/2106.10304}{{\tt 2106.10304}}].

\bibitem{Frankel}
N.~{Frankel}, A.~{Pillepich}, H.-W. {Rix}, V.~{Rodriguez-Gomez}, J.~{Sanders},
  J.~{Bovy} et~al., \emph{{Simulated Bars May Be Shorter But Are Not Slower
  Than Observed: TNG50 vs. MaNGA}}, {\emph{arXiv e-prints} (Jan., 2022)
  arXiv:2201.08406}, [\href{https://arxiv.org/abs/2201.08406}{{\tt
  2201.08406}}].

\bibitem{Fragkoudi}
F.~{Fragkoudi}, R.~J.~J. {Grand}, R.~{Pakmor}, V.~{Springel}, S.~D.~M. {White},
  F.~{Marinacci} et~al., \emph{{Revisiting the tension between fast bars and
  the {\ensuremath{\Lambda}}CDM paradigm}},
  \href{http://dx.doi.org/10.1051/0004-6361/202140320}{\emph{\aap} {\bf 650}
  (June, 2021) L16}, [\href{https://arxiv.org/abs/2011.13942}{{\tt
  2011.13942}}].

\bibitem{Rey:2018cfb}
M.~P. Rey, A.~Pontzen and A.~Saintonge, \emph{{Sensitivity of dark matter
  haloes to their accretion histories}},
  \href{http://dx.doi.org/10.1093/mnras/stz552}{\emph{Mon. Not. Roy. Astron.
  Soc.} {\bf 485} (2019) 1906--1915},
  [\href{https://arxiv.org/abs/1810.09473}{{\tt 1810.09473}}].

\bibitem{Stopyra:2020egb}
S.~Stopyra, A.~Pontzen, H.~Peiris, N.~Roth and M.~Rey, \emph{{GenetIC -- a new
  initial conditions generator to support genetically modified zoom
  simulations}},
  \href{http://dx.doi.org/10.3847/1538-4365/abcd94}{\emph{Astrophys. J. Suppl.}
  {\bf 252} (2021) 28}, [\href{https://arxiv.org/abs/2006.01841}{{\tt
  2006.01841}}].

\bibitem{Cadiou:2021chs}
C.~Cadiou, A.~Pontzen, H.~V. Peiris and L.~Lucie-Smith, \emph{{The causal
  effect of environment on halo mass and concentration}},
  \href{http://dx.doi.org/10.1093/mnras/stab2650}{\emph{Mon. Not. Roy. Astron.
  Soc.} {\bf 508} (2021) 1189--1194},
  [\href{https://arxiv.org/abs/2107.03407}{{\tt 2107.03407}}].

\bibitem{Cadiou:2022krq}
C.~Cadiou, A.~Pontzen and H.~V. Peiris, \emph{{Stellar angular momentum can be
  controlled from cosmological initial conditions}},
  \href{https://arxiv.org/abs/2206.11913}{{\tt 2206.11913}}.

\bibitem{Avila-Reese:2003cjm}
V.~Avila-Reese, P.~Colin, G.~Piccinelli and C.~Firmani, \emph{{The effects of
  non-Gaussian initial conditions on the structure and substructure of cold
  dark matter halos}}, \href{http://dx.doi.org/10.1086/378773}{\emph{Astrophys.
  J.} {\bf 598} (2003) 36--48},
  [\href{https://arxiv.org/abs/astro-ph/0306293}{{\tt astro-ph/0306293}}].

\bibitem{Dalal:2007cu}
N.~Dalal, O.~Dore, D.~Huterer and A.~Shirokov, \emph{{The imprints of
  primordial non-gaussianities on large-scale structure: scale dependent bias
  and abundance of virialized objects}},
  \href{http://dx.doi.org/10.1103/PhysRevD.77.123514}{\emph{Phys. Rev. D} {\bf
  77} (2008) 123514}, [\href{https://arxiv.org/abs/0710.4560}{{\tt
  0710.4560}}].

\bibitem{LoVerde:2007ri}
M.~LoVerde, A.~Miller, S.~Shandera and L.~Verde, \emph{{Effects of
  Scale-Dependent Non-Gaussianity on Cosmological Structures}},
  \href{http://dx.doi.org/10.1088/1475-7516/2008/04/014}{\emph{JCAP} {\bf 04}
  (2008) 014}, [\href{https://arxiv.org/abs/0711.4126}{{\tt 0711.4126}}].

\bibitem{Pillepich:2008ka}
A.~Pillepich, C.~Porciani and O.~Hahn, \emph{{Universal halo mass function and
  scale-dependent bias from N-body simulations with non-Gaussian initial
  conditions}},
  \href{http://dx.doi.org/10.1111/j.1365-2966.2009.15914.x}{\emph{Mon. Not.
  Roy. Astron. Soc.} {\bf 402} (2010) 191--206},
  [\href{https://arxiv.org/abs/0811.4176}{{\tt 0811.4176}}].

\bibitem{Smith:2010fh}
R.~E. Smith, V.~Desjacques and L.~Marian, \emph{{Nonlinear clustering in models
  with primordial non-Gaussianity: the halo model approach}},
  \href{http://dx.doi.org/10.1103/PhysRevD.83.043526}{\emph{Phys. Rev. D} {\bf
  83} (2011) 043526}, [\href{https://arxiv.org/abs/1009.5085}{{\tt
  1009.5085}}].

\bibitem{MoradinezhadDizgah:2013rkr}
A.~Moradinezhad~Dizgah, S.~Dodelson and A.~Riotto, \emph{{Imprint of Primordial
  Non-Gaussianity on Dark Matter Halo Profiles}},
  \href{http://dx.doi.org/10.1103/PhysRevD.88.063513}{\emph{Phys. Rev. D} {\bf
  88} (2013) 063513}, [\href{https://arxiv.org/abs/1307.2632}{{\tt
  1307.2632}}].

\bibitem{Springel:2020plp}
V.~Springel, R.~Pakmor, O.~Zier and M.~Reinecke, \emph{{Simulating cosmic
  structure formation with the gadget-4 code}},
  \href{http://dx.doi.org/10.1093/mnras/stab1855}{\emph{Mon. Not. Roy. Astron.
  Soc.} {\bf 506} (2021) 2871--2949},
  [\href{https://arxiv.org/abs/2010.03567}{{\tt 2010.03567}}].

\bibitem{Maldacena:2002vr}
J.~M. Maldacena, \emph{{Non-Gaussian features of primordial fluctuations in
  single field inflationary models}},
  \href{http://dx.doi.org/10.1088/1126-6708/2003/05/013}{\emph{JHEP} {\bf 05}
  (2003) 013}, [\href{https://arxiv.org/abs/astro-ph/0210603}{{\tt
  astro-ph/0210603}}].

\bibitem{Creminelli:2004yq}
P.~Creminelli and M.~Zaldarriaga, \emph{{Single field consistency relation for
  the 3-point function}},
  \href{http://dx.doi.org/10.1088/1475-7516/2004/10/006}{\emph{JCAP} {\bf 10}
  (2004) 006}, [\href{https://arxiv.org/abs/astro-ph/0407059}{{\tt
  astro-ph/0407059}}].

\bibitem{Biagetti:2019bnp}
M.~Biagetti, \emph{{The Hunt for Primordial Interactions in the Large Scale
  Structures of the Universe}},
  \href{http://dx.doi.org/10.3390/galaxies7030071}{\emph{Galaxies} {\bf 7}
  (2019) 71}, [\href{https://arxiv.org/abs/1906.12244}{{\tt 1906.12244}}].

\bibitem{Planck:2019kim}
{\scshape Planck} collaboration, Y.~Akrami et~al., \emph{{Planck 2018 results.
  IX. Constraints on primordial non-Gaussianity}},
  \href{http://dx.doi.org/10.1051/0004-6361/201935891}{\emph{Astron.
  Astrophys.} {\bf 641} (2020) A9},
  [\href{https://arxiv.org/abs/1905.05697}{{\tt 1905.05697}}].

\bibitem{Mueller:2021tqa}
E.-M. Mueller et~al., \emph{{The clustering of galaxies in the completed
  SDSS-IV extended Baryon Oscillation Spectroscopic Survey: Primordial
  non-Gaussianity in Fourier Space}},
  \href{https://arxiv.org/abs/2106.13725}{{\tt 2106.13725}}.

\bibitem{Cabass:2022ymb}
G.~Cabass, M.~M. Ivanov, O.~H.~E. Philcox, M.~Simonovi\'c and M.~Zaldarriaga,
  \emph{{Constraints on multifield inflation from the BOSS galaxy survey}},
  \href{http://dx.doi.org/10.1103/PhysRevD.106.043506}{\emph{Phys. Rev. D} {\bf
  106} (2022) 043506}, [\href{https://arxiv.org/abs/2204.01781}{{\tt
  2204.01781}}].

\bibitem{Chevallard:2014sxa}
J.~Chevallard, J.~Silk, T.~Nishimichi, M.~Habouzit, G.~A. Mamon and S.~Peirani,
  \emph{{Effect of primordial non-Gaussianities on the far-UV luminosity
  function of high-redshift galaxies: implications for cosmic reionization}},
  \href{http://dx.doi.org/10.1093/mnras/stu2280}{\emph{Mon. Not. Roy. Astron.
  Soc.} {\bf 446} (2015) 3235--3252},
  [\href{https://arxiv.org/abs/1410.7768}{{\tt 1410.7768}}].

\bibitem{Khoury:2008wj}
J.~Khoury and F.~Piazza, \emph{{Rapidly-Varying Speed of Sound, Scale
  Invariance and Non-Gaussian Signatures}},
  \href{http://dx.doi.org/10.1088/1475-7516/2009/07/026}{\emph{JCAP} {\bf 07}
  (2009) 026}, [\href{https://arxiv.org/abs/0811.3633}{{\tt 0811.3633}}].

\bibitem{Riotto:2010nh}
A.~Riotto and M.~S. Sloth, \emph{{Strongly Scale-dependent Non-Gaussianity}},
  \href{http://dx.doi.org/10.1103/PhysRevD.83.041301}{\emph{Phys. Rev. D} {\bf
  83} (2011) 041301}, [\href{https://arxiv.org/abs/1009.3020}{{\tt
  1009.3020}}].

\bibitem{Byrnes:2011gh}
C.~T. Byrnes, K.~Enqvist, S.~Nurmi and T.~Takahashi, \emph{{Strongly
  scale-dependent polyspectra from curvaton self-interactions}},
  \href{http://dx.doi.org/10.1088/1475-7516/2011/11/011}{\emph{JCAP} {\bf 11}
  (2011) 011}, [\href{https://arxiv.org/abs/1108.2708}{{\tt 1108.2708}}].

\bibitem{Khatri:2015tla}
R.~Khatri and R.~Sunyaev, \emph{{Constraints on \ensuremath{\mu}-distortion
  fluctuations and primordial non-Gaussianity from Planck data}},
  \href{http://dx.doi.org/10.1088/1475-7516/2015/9/026}{\emph{JCAP} {\bf 09}
  (2015) 026}, [\href{https://arxiv.org/abs/1507.05615}{{\tt 1507.05615}}].

\bibitem{Sabti:2020ser}
N.~Sabti, J.~B. Mu\~noz and D.~Blas, \emph{{First Constraints on Small-Scale
  Non-Gaussianity from UV Galaxy Luminosity Functions}},
  \href{http://dx.doi.org/10.1088/1475-7516/2021/01/010}{\emph{JCAP} {\bf 01}
  (2021) 010}, [\href{https://arxiv.org/abs/2009.01245}{{\tt 2009.01245}}].

\bibitem{Rotti:2022lvy}
A.~Rotti, A.~Ravenni and J.~Chluba, \emph{{Non-Gaussianity constraints with
  anisotropic \ensuremath{\mu} distortion measurements from Planck}},
  \href{http://dx.doi.org/10.1093/mnras/stac2082}{\emph{Mon. Not. Roy. Astron.
  Soc.} {\bf 515} (2022) 5847--5868},
  [\href{https://arxiv.org/abs/2205.15971}{{\tt 2205.15971}}].

\bibitem{Bianchini:2022dqh}
F.~Bianchini and G.~Fabbian, \emph{{CMB spectral distortions revisited: A new
  take on \ensuremath{<}math
  display=''inline''\ensuremath{>}\ensuremath{<}mi\ensuremath{>}\ensuremath{\mu}\ensuremath{<}/mi\ensuremath{>}\ensuremath{<}/math\ensuremath{>}
  distortions and primordial non-Gaussianities from FIRAS data}},
  \href{http://dx.doi.org/10.1103/PhysRevD.106.063527}{\emph{Phys. Rev. D} {\bf
  106} (2022) 063527}, [\href{https://arxiv.org/abs/2206.02762}{{\tt
  2206.02762}}].

\bibitem{Blas:2011rf}
D.~Blas, J.~Lesgourgues and T.~Tram, \emph{{The Cosmic Linear Anisotropy
  Solving System (CLASS) II: Approximation schemes}},
  \href{http://dx.doi.org/10.1088/1475-7516/2011/07/034}{\emph{JCAP} {\bf 07}
  (2011) 034}, [\href{https://arxiv.org/abs/1104.2933}{{\tt 1104.2933}}].

\bibitem{Bucher:2015ura}
M.~Bucher, B.~Racine and B.~van Tent, \emph{{The binned bispectrum estimator:
  template-based and non-parametric CMB non-Gaussianity searches}},
  \href{http://dx.doi.org/10.1088/1475-7516/2016/05/055}{\emph{JCAP} {\bf 05}
  (2016) 055}, [\href{https://arxiv.org/abs/1509.08107}{{\tt 1509.08107}}].

\bibitem{Rezaie:2021voi}
M.~Rezaie et~al., \emph{{Primordial non-Gaussianity from the completed SDSS-IV
  extended Baryon Oscillation Spectroscopic Survey \textendash{} I: Catalogue
  preparation and systematic mitigation}},
  \href{http://dx.doi.org/10.1093/mnras/stab1730}{\emph{Mon. Not. Roy. Astron.
  Soc.} {\bf 506} (2021) 3439--3454},
  [\href{https://arxiv.org/abs/2106.13724}{{\tt 2106.13724}}].

\bibitem{Hahn:2011uy}
O.~Hahn and T.~Abel, \emph{{Multi-scale initial conditions for cosmological
  simulations}},
  \href{http://dx.doi.org/10.1111/j.1365-2966.2011.18820.x}{\emph{Mon. Not.
  Roy. Astron. Soc.} {\bf 415} (2011) 2101--2121},
  [\href{https://arxiv.org/abs/1103.6031}{{\tt 1103.6031}}].

\bibitem{Michaux:2020yis}
M.~Michaux, O.~Hahn, C.~Rampf and R.~E. Angulo, \emph{{Accurate initial
  conditions for cosmological N-body simulations: Minimizing truncation and
  discreteness errors}},
  \href{http://dx.doi.org/10.1093/mnras/staa3149}{\emph{Mon. Not. Roy. Astron.
  Soc.} {\bf 500} (2020) 663--683},
  [\href{https://arxiv.org/abs/2008.09588}{{\tt 2008.09588}}].

\bibitem{Springel:2000qu}
V.~Springel, S.~D.~M. White, G.~Tormen and G.~Kauffmann, \emph{{Populating a
  cluster of galaxies. 1. Results at z = 0}},
  \href{http://dx.doi.org/10.1046/j.1365-8711.2001.04912.x}{\emph{Mon. Not.
  Roy. Astron. Soc.} {\bf 328} (2001) 726},
  [\href{https://arxiv.org/abs/astro-ph/0012055}{{\tt astro-ph/0012055}}].

\bibitem{Fakhouri10}
O.~{Fakhouri}, C.-P. {Ma} and M.~{Boylan-Kolchin}, \emph{{The merger rates and
  mass assembly histories of dark matter haloes in the two Millennium
  simulations}},
  \href{http://dx.doi.org/10.1111/j.1365-2966.2010.16859.x}{\emph{\mnras} {\bf
  406} (Aug., 2010) 2267--2278}, [\href{https://arxiv.org/abs/1001.2304}{{\tt
  1001.2304}}].

\bibitem{Amon:2022azi}
A.~Amon and G.~Efstathiou, \emph{{A non-linear solution to the $S_8$
  tension?}},  \href{https://arxiv.org/abs/2206.11794}{{\tt 2206.11794}}.

\bibitem{ytree}
B.~D. Smith and M.~Lang, \emph{ytree: A python package for analyzing merger
  trees}, \href{http://dx.doi.org/10.21105/joss.01881}{\emph{Journal of Open
  Source Software} {\bf 4} (dec, 2019) 1881}.

\bibitem{Fakhouri:2008cn}
O.~Fakhouri and C.-P. Ma, \emph{{Environmental Dependence of Dark Matter Halo
  Growth I: Halo Merger Rates}},
  \href{http://dx.doi.org/10.1111/j.1365-2966.2009.14480.x}{\emph{Mon. Not.
  Roy. Astron. Soc.} {\bf 394} (2009) 1825},
  [\href{https://arxiv.org/abs/0808.2471}{{\tt 0808.2471}}].

\bibitem{Biagetti:2022ode}
M.~Biagetti, G.~Franciolini and A.~Riotto, \emph{{The JWST High Redshift
  Observations and Primordial Non-Gaussianity}},
  \href{https://arxiv.org/abs/2210.04812}{{\tt 2210.04812}}.

\bibitem{Labbe}
I.~{Labbe}, P.~{van Dokkum}, E.~{Nelson}, R.~{Bezanson}, K.~{Suess}, J.~{Leja}
  et~al., \emph{{A very early onset of massive galaxy formation}}, {\emph{arXiv
  e-prints} (July, 2022) arXiv:2207.12446},
  [\href{https://arxiv.org/abs/2207.12446}{{\tt 2207.12446}}].

\bibitem{MBK2}
M.~{Boylan-Kolchin}, \emph{{Stress Testing $\Lambda$CDM with High-redshift
  Galaxy Candidates}}, {\emph{arXiv e-prints} (Aug., 2022) arXiv:2208.01611},
  [\href{https://arxiv.org/abs/2208.01611}{{\tt 2208.01611}}].

\bibitem{Dalal:2008zd}
N.~Dalal, M.~White, J.~R. Bond and A.~Shirokov, \emph{{Halo Assembly Bias in
  Hierarchical Structure Formation}},
  \href{http://dx.doi.org/10.1086/591512}{\emph{Astrophys. J.} {\bf 687} (2008)
  12--21}, [\href{https://arxiv.org/abs/0803.3453}{{\tt 0803.3453}}].

\bibitem{Lazeyras:2022koc}
T.~Lazeyras, A.~Barreira, F.~Schmidt and V.~Desjacques, \emph{{Assembly bias in
  the local PNG halo bias and its implication for $f_{\rm NL}$ constraints}},
  \href{https://arxiv.org/abs/2209.07251}{{\tt 2209.07251}}.

\bibitem{Pawlowski2}
M.~S. {Pawlowski}, \emph{{Phase-Space Correlations among Systems of Satellite
  Galaxies}}, \href{http://dx.doi.org/10.3390/galaxies9030066}{\emph{Galaxies}
  {\bf 9} (Sept., 2021) 66}, [\href{https://arxiv.org/abs/2109.02654}{{\tt
  2109.02654}}].

\bibitem{LyndenBell}
D.~{Lynden-Bell}, \emph{{Dwarf galaxies and globular clusters in high velocity
  hydrogen streams.}},
  \href{http://dx.doi.org/10.1093/mnras/174.3.695}{\emph{\mnras} {\bf 174}
  (Mar., 1976) 695--710}.

\bibitem{Pawlowski3}
M.~S. {Pawlowski}, J.~{Pflamm-Altenburg} and P.~{Kroupa}, \emph{{The VPOS: a
  vast polar structure of satellite galaxies, globular clusters and streams
  around the Milky Way}},
  \href{http://dx.doi.org/10.1111/j.1365-2966.2012.20937.x}{\emph{\mnras} {\bf
  423} (June, 2012) 1109--1126}, [\href{https://arxiv.org/abs/1204.5176}{{\tt
  1204.5176}}].

\bibitem{Wang:2012bt}
J.~Wang, C.~S. Frenk and A.~P. Cooper, \emph{{The Spatial Distribution of
  Galactic Satellites in the LCDM Cosmology}},
  \href{http://dx.doi.org/10.1093/mnras/sts442}{\emph{Mon. Not. Roy. Astron.
  Soc.} {\bf 429} (2013) 1502}, [\href{https://arxiv.org/abs/1206.1340}{{\tt
  1206.1340}}].

\bibitem{LiH}
H.~{Li}, F.~{Hammer}, C.~{Babusiaux}, M.~S. {Pawlowski}, Y.~{Yang}, F.~{Arenou}
  et~al., \emph{{Gaia EDR3 Proper Motions of Milky Way Dwarfs. I. 3D Motions
  and Orbits}}, \href{http://dx.doi.org/10.3847/1538-4357/ac0436}{\emph{\apj}
  {\bf 916} (July, 2021) 8}, [\href{https://arxiv.org/abs/2104.03974}{{\tt
  2104.03974}}].

\bibitem{Sawala:2022xom}
T.~Sawala, M.~Cautun, C.~S. Frenk, J.~Helly, J.~Jasche, A.~Jenkins et~al.,
  \emph{{The Milky Way's plane of satellites: consistent with $\Lambda$CDM}},
  \href{https://arxiv.org/abs/2205.02860}{{\tt 2205.02860}}.

\bibitem{IbataM31}
R.~A. {Ibata}, G.~F. {Lewis}, A.~R. {Conn}, M.~J. {Irwin}, A.~W. {McConnachie},
  S.~C. {Chapman} et~al., \emph{{A vast, thin plane of corotating dwarf
  galaxies orbiting the Andromeda galaxy}},
  \href{http://dx.doi.org/10.1038/nature11717}{\emph{\nat} {\bf 493} (Jan.,
  2013) 62--65}, [\href{https://arxiv.org/abs/1301.0446}{{\tt 1301.0446}}].

\bibitem{Cameron:2010bh}
E.~Cameron, \emph{{On the Estimation of Confidence Intervals for Binomial
  Population Proportions in Astronomy: The Simplicity and Superiority of the
  Bayesian Approach}}, \href{http://dx.doi.org/10.1071/AS10046}{\emph{Publ.
  Astron. Soc. Austral.} {\bf 28} (2011) 128},
  [\href{https://arxiv.org/abs/1012.0566}{{\tt 1012.0566}}].

\bibitem{Dalal11}
S.~{Shandera}, N.~{Dalal} and D.~{Huterer}, \emph{{A generalized local ansatz
  and its effect on halo bias}},
  \href{http://dx.doi.org/10.1088/1475-7516/2011/03/017}{\emph{\jcap} {\bf
  2011} (Mar., 2011) 017}, [\href{https://arxiv.org/abs/1010.3722}{{\tt
  1010.3722}}].

\bibitem{Pena:2022sdg}
G.~A. Pe\~na and G.~N. Candlish, \emph{{The large-scale structure from
  non-Gaussian primordial perturbations}},
  \href{http://dx.doi.org/10.1093/mnras/stac206}{\emph{Mon. Not. Roy. Astron.
  Soc.} {\bf 511} (2022) 2259--2273},
  [\href{https://arxiv.org/abs/2201.08842}{{\tt 2201.08842}}].

\bibitem{stahlDDM}
C.~{Stahl}, B.~{Famaey}, G.~{Thomas}, Y.~{Dubois} and R.~{Ibata},
  \emph{{Dipolar dark matter simulations on galaxy scales with the RAMSES
  code}}, {\emph{arXiv e-prints} (Sept., 2022) arXiv:2209.07831},
  [\href{https://arxiv.org/abs/2209.07831}{{\tt 2209.07831}}].

\bibitem{Murgia:2017lwo}
R.~Murgia, A.~Merle, M.~Viel, M.~Totzauer and A.~Schneider, \emph{{''Non-cold''
  dark matter at small scales: a general approach}},
  \href{http://dx.doi.org/10.1088/1475-7516/2017/11/046}{\emph{JCAP} {\bf 11}
  (2017) 046}, [\href{https://arxiv.org/abs/1704.07838}{{\tt 1704.07838}}].

\bibitem{Turk:2010ah}
M.~J. {Turk}, B.~D. {Smith}, J.~S. {Oishi}, S.~{Skory}, S.~W. {Skillman},
  T.~{Abel} et~al., \emph{{yt: A Multi-code Analysis Toolkit for Astrophysical
  Simulation Data}},
  \href{http://dx.doi.org/10.1088/0067-0049/192/1/9}{\emph{\apjs} {\bf 192}
  (Jan., 2011) 9}, [\href{https://arxiv.org/abs/1011.3514}{{\tt 1011.3514}}].

\bibitem{Perez:2007emg}
F.~Perez and B.~E. Granger, \emph{{IPython: A System for Interactive Scientific
  Computing}}, \href{http://dx.doi.org/10.1109/MCSE.2007.53}{\emph{Comput. Sci.
  Eng.} {\bf 9} (2007) 21--29}.

\bibitem{Hunter:2007ouj}
J.~D. Hunter, \emph{{Matplotlib: A 2D Graphics Environment}},
  \href{http://dx.doi.org/10.1109/MCSE.2007.55}{\emph{Comput. Sci. Eng.} {\bf
  9} (2007) 90--95}.

\bibitem{vanderWalt:2011bqk}
S.~van~der Walt, S.~C. Colbert and G.~Varoquaux, \emph{{The NumPy Array: A
  Structure for Efficient Numerical Computation}},
  \href{http://dx.doi.org/10.1109/MCSE.2011.37}{\emph{Comput. Sci. Eng.} {\bf
  13} (2011) 22--30}, [\href{https://arxiv.org/abs/1102.1523}{{\tt
  1102.1523}}].

\bibitem{berthoud}
F.~Berthoud, B.~Bzeznik, N.~Gibelin, M.~Laurens, C.~Bonamy, M.~Morel et~al.,
  \emph{{Estimation de l'empreinte carbone d'une heure.coeur de calcul}},
  research report, {UGA - Universit{\'e} Grenoble Alpes ; CNRS ; INP Grenoble ;
  INRIA}, Apr., 2020.

\end{thebibliography}\endgroup

\end{document}